\documentclass[aps,prx,twocolumn,nopacs,superscriptaddress,floatfix]{revtex4-2}

\usepackage[breaklinks=true,colorlinks,citecolor=blue,linkcolor=blue,urlcolor=blue]{hyperref}
\usepackage{epsfig,mathrsfs,xcolor,latexsym,subfigure,marginnote,graphicx,verbatim,relsize,mathrsfs,color,array,amsmath,amsfonts,amssymb,graphicx}
\usepackage{xspace}
\usepackage{placeins}

\newcommand{\tss}[1]{\ensuremath{^{\text{#1}}}}
\newcommand{\rrscan}{\(r\)\tss{2}SCAN\xspace}

\newcommand{\SB}[1]{{\it #1}}
\newcommand{\veck}{\boldsymbol{k}}
\newcommand\vecq{\boldsymbol{q}}

\usepackage{booktabs}   %
\usepackage{multirow}   %
\usepackage[referable]{threeparttablex}
\newcolumntype{x}{>{$}l<{$}}  %
\newcolumntype{y}{>{$}c<{$}}  %
\newcolumntype{z}{>{$}r<{$}}  %

\usepackage{braket}
\usepackage{textcase}
\usepackage{makecell}

\newcommand{\SI}[2]{\ensuremath{#1\,\mathrm{#2}}}
\newcommand{\si}[1]{\ensuremath{\mathrm{#1}}}
\newcommand{\angstrom}{\text{\AA}}
\newcommand{\beginsupplement}{%
  \setcounter{equation}{0}
  \renewcommand{\theequation}{S\arabic{equation}}%
  \setcounter{table}{0}
  \renewcommand{\thetable}{S\arabic{table}}%
  \setcounter{figure}{0}
  \renewcommand{\thefigure}{S\arabic{figure}}%
  \setcounter{section}{0}
  \renewcommand{\thesection}{S-\Roman{section}}%
  \setcounter{subsection}{0}
  \renewcommand{\thesubsection}{S-\Roman{section}.\Alph{subsection}}%
  \renewcommand{\theHequation}{S\arabic{equation}}%
  \renewcommand{\theHtable}{S\arabic{table}}%
  \renewcommand{\theHfigure}{S\arabic{figure}}%
  \renewcommand{\theHsection}{S\arabic{section}}%
  \renewcommand{\theHsubsection}{S\arabic{section}.\arabic{subsection}}%
}

\makeatletter
\let\author@extramark\@empty
\newcommand\sharedmark[1]{\gdef\author@extramark{\comma@space\frontmatter@footnotemark{#1}}}
\def\doauthor#1#2#3{%
  \ignorespaces#1\unskip\@listcomma
  \begingroup
   #3\author@extramark\global\let\author@extramark\@empty
  \@if@empty{#2}{\endgroup{}{}}{\endgroup{\comma@space}{}\frontmatter@footnote{#2}}%
  \space \@listand
}%
\makeatother

\begin{document}

\title{Rotational Phonons Activated by Magnetism as a Key Ingredient of Cuprate Physics}

\author{Yanyong~Wang}
\thanks{These authors contributed equally to this work}
\affiliation{Department of Physics and Engineering Physics, Tulane University, New Orleans, LA 70118, USA}

\author{Manuel~Engel}
\thanks{These authors contributed equally to this work}
\affiliation{VASP Software GmbH, Berggasse 21/14, 1090 Vienna, Austria}

\author{Christopher Lane}
\affiliation{Theoretical Division, Los Alamos National Laboratory, Los Alamos, New Mexico 87545, USA}

\author{{Sougata Mardanya}}
\affiliation{Department of Physics and Astronomy, Howard University, Washington DC 20059, USA}

\author{{Bahadur Singh}}
\affiliation{Department of Condensed Matter Physics and Materials Science, Tata Institute of Fundamental Research, Mumbai 400005, India}

\author{Henrique Miranda}
\affiliation{VASP Software GmbH, Berggasse 21/14, 1090 Vienna, Austria}

\author{Lin Hou}
\affiliation{Department of Physics and Engineering Physics, Tulane University, New Orleans, LA 70118, USA}

\author{{Sugata Chowdhury}}
\affiliation{Department of Physics and Astronomy, Howard University, Washington DC 20059, USA}

\author{Bernardo~Barbiellini}
\affiliation{Department of Physics, School of Engineering Science, LUT University, FI-53850 Lappeenranta, Finland}
\affiliation{Department of Physics, Northeastern University, Boston, MA 02115, USA}
\affiliation{Quantum Materials and Sensing Institute, Northeastern University, Burlington, MA 01803, USA}

\author{Adrienn Ruzsinszky}
\affiliation{Department of Physics and Engineering Physics, Tulane University, New Orleans, LA 70118, USA}

\author{John P. Perdew}
\affiliation{Department of Physics and Engineering Physics, Tulane University, New Orleans, LA 70118, USA}

\author{Robert S. Markiewicz}
\affiliation{Department of Physics, Northeastern University, Boston, MA 02115, USA}
\affiliation{Quantum Materials and Sensing Institute, Northeastern University, Burlington, MA 01803, USA}

\author{Arun~Bansil}
\email[Corresponding author:~]{ar.bansil@neu.edu}
\affiliation{Department of Physics, Northeastern University, Boston, MA 02115, USA}
\affiliation{Quantum Materials and Sensing Institute, Northeastern University, Burlington, MA 01803, USA}

\author{Jianwei Sun}
\email[Corresponding author:~]{jsun@tulane.edu}
\affiliation{Department of Physics and Engineering Physics, Tulane University, New Orleans, LA 70118, USA}

\author{Ruiqi~Zhang\sharedmark{1}}
\email[Corresponding author:~]{rzhang16@tulane.edu}
\affiliation{Department of Physics and Engineering Physics, Tulane University, New Orleans, LA 70118, USA}

\begin{abstract}

Circularly polarized lattice vibrations carry angular momentum and can link lattice dynamics to collective responses in strongly correlated materials, particularly cuprate superconductors. Angle-resolved photoemission spectroscopy (ARPES) reveals ubiquitous quasiparticle ``kinks'' near $\sim$70 and $\sim$40 meV in cuprates, often accompanied by peak--dip--hump (PDH) features, indicating strong coupling between electrons and low-energy bosonic excitations. However, the relevant phonon modes and the role of circularly polarized vibrations in these spectral anomalies remain largely elusive. Here, we investigate electron--phonon coupling (EPC) in hole-doped infinite-layer CaCuO$_2$ at four selective hole dopings (12.5\%--56.25\%) using the Strongly Constrained and Appropriately Normed (SCAN) density functional with explicit magnetic interactions. At 25\% doping, the magnetic phase exhibits substantially stronger EPC ($\lambda \sim 0.5$), five times that of the nonmagnetic phase, producing kinks and PDH features over 40--80 meV, with self-energy magnitudes comparable to experiment. Oxygen rotational phonons dominate the coupling over 50--65~meV, while breathing modes play a lesser role; the rotational modes emerge exclusively in the magnetic phase. {At 37.5\% doping, $\lambda$ increases to $\sim$1.6 due to strong coupling of low-frequency phonons near the Fermi-surface nesting wave vector, whereas at 56.25\%, where Cu magnetic moments vanish, $\lambda$ decreases to $\sim$0.4. The rotational modes carry staggered angular momentum that cancels in the unperturbed lattice; a time-reversal-breaking gyrotropic coupling on Cu sites lifts this cancellation, yielding a finite phonon thermal Hall conductivity.} Our findings establish magnetism-enhanced EPC as an important ingredient of cuprate physics and identify rotational oxygen phonons as relevant to spectral anomalies, superconductivity, and the pseudogap thermal Hall effect.

\end{abstract} 

\maketitle

\section{Introduction}

Despite nearly four decades of extensive research, the pairing mechanism responsible for high-temperature superconductivity in cuprates remains elusive~\cite{Shen_ARPS_cuprate,AR_QM,Linear_cuprate,Norman_science}. This materials family is widely regarded as an archetype of strongly correlated systems, in which superconductivity emerges upon doping an antiferromagnetic Mott or charge-transfer insulating parent compound. The resulting phase diagram hosts a wealth of correlation-driven phenomena, including predominantly $d$-wave pairing, strange-metal transport characterized by a near-linear temperature dependence of the resistivity near optimal doping, unusually large magnetic exchange interactions, and intertwined electronic orders~\cite{Kievelson2003,Eduardo2015,Elbio2005,Emery1999PNAS,Hanaguri2004nature,Kohsaka2007a,Pan2018nature,Jiang2022PNAS,Tranquada1996,Zheng2017,Huang2017a}.

Although the phenomenology of cuprates has motivated many theories emphasizing purely electronic mechanisms, a growing body of experimental results has indicated that lattice degrees of freedom play a key role. Momentum-resolved scattering experiments have revealed pronounced phonon anomalies that track intertwined orders such as stripes and charge-density waves~\cite{Ding_PNAS}, and spectroscopic probes report broad bosonic signatures together with measurable oxygen-isotope effects~\cite{Lee2006}. Angle-resolved photoemission spectroscopy (ARPES) reveals a rapid co-evolution of superconductivity and electron--phonon coupling (EPC) in Bi-2212 (Bi$_2$Sr$_2$CaCu$_2$O$_{8+\delta}$)~\cite{He_Rapid}, and a large body of ARPES studies shows that low-energy electronic excitations near the Fermi level couple strongly to bosonic modes to drive quasiparticle ``kinks'' across the cuprate family~\cite{ShenZX_PRL_nodal70,Johnson_PRL_70,Kaminski_PRL_70,ShenZX_Nature_70,Zhou2003_Nature_nodal,Gromko_PRB_70,Sato_PRL_nodal_antinodal,Kordyuk_nodal,Zhou_PRL_nodal,JunFeng_Hu_nodal_antinodal,Kondo_nodal_antinodal,Anzai2017_nodal_antinodal,ZXShen_PRL_40,Kim_antinodal_40,ShenZX_B1g}.

The kinks in cuprates were long classified into two characteristic energy scales: a nodal kink near 70~meV~\cite{ShenZX_PRL_nodal70,Johnson_PRL_70,Kaminski_PRL_70,ShenZX_Nature_70,Zhou2003_Nature_nodal,Gromko_PRB_70,Sato_PRL_nodal_antinodal,Kordyuk_nodal,Zhou_PRL_nodal,JunFeng_Hu_nodal_antinodal,Kondo_nodal_antinodal,Anzai2017_nodal_antinodal} and an antinodal kink near 40~meV~\cite{ZXShen_PRL_40,Gromko_PRB_70,Kim_antinodal_40,Sato_PRL_nodal_antinodal,ShenZX_B1g,JunFeng_Hu_nodal_antinodal,Kondo_nodal_antinodal,Anzai2017_nodal_antinodal}. The microscopic origin of these kinks remains a subject of active debate. Both phonons~\cite{ShenZX_Nature_70,ShenZX_PRL_nodal70,ZXShen_PRL_40} and spin fluctuations~\cite{Johnson_PRL_70,Byczuk2007,Dahm2009,norman_prl_sf} have been proposed as the relevant bosonic excitations. Recent ultrahigh-resolution ARPES measurements report both the nodal and antinodal kinks to be insensitive to temperature and doping~\cite{Hongtao_PNAS}. This robustness does not favor a purely magnetic origin and points to EPC as a key player in kink physics. Taken together, these observations highlight a central role for EPC in cuprates, alongside strong electronic correlations. Beyond spectral anomalies, the large thermal Hall effect in the pseudogap phase~\cite{TaillThPh} implicates chiral phonons as critical heat carriers, yet the microscopic identity of these modes in cuprates remains unknown~\cite{Ramshaw2026}.

From a theoretical perspective, the role of EPC in cuprates has long been controversial. Early density-functional theory (DFT) calculations concluded that EPC is too weak to account for the experimentally observed kinks~\cite{giustino2008cuprate}. However, this conclusion has been questioned~\cite{reznik2008photoemission,ZhengluGWPTCuprate}, as conventional DFT approximations---such as the local density approximation (LDA) and the Perdew--Burke--Ernzerhof (PBE) generalized gradient approximation---often fail to capture the electronic structure, phonon properties, and correlated magnetism of cuprates~\cite{Furness2018,Lane2018,ZhangYPANS,NingPRBYBCO6,NingJCPDFAYBCO6}. More recently, GW perturbation theory (GWPT) based calculations have revealed a correlation-enhanced EPC with (dimensionless) strengths $\lambda$ that are two to three times greater than those obtained from standard DFT~\cite{ZhengluGWPTCuprate}. Both early DFT~\cite{giustino2008cuprate} and GWPT~\cite{ZhengluGWPTCuprate} calculations are typically based on nonmagnetic reference states that neglect magnetic effects, despite growing evidence that magnetism and spin fluctuations are intrinsic to cuprate physics~\cite{Tranquada2004}. These considerations motivate a material-specific assessment of EPC in cuprates that includes magnetism explicitly. Such an approach is needed to identify the phonon modes responsible for the ARPES kinks and to assess whether magnetism qualitatively reshapes the electron--phonon interaction. In particular, recent work has demonstrated that magnetism can induce circularly polarized phonon modes~\cite{Che_PRL_2025,Yang_NP_2026}, raising the question of whether magnetism activates such rotational modes in cuprates and whether these modes underlie the observed kinks and the pseudogap thermal Hall effect, and how the EPC evolves with doping.

We address this challenge by investigating EPC in the exemplar hole-doped infinite-layer cuprate CaCuO$_2$ using the Strongly Constrained and Appropriately Normed (SCAN) density functional~\cite{Sun2015}, with magnetic effects included explicitly. The SCAN density functional employed here is computationally semilocal but employs the positive orbital kinetic energy density, a fully nonlocal functional of the electron density. It satisfies 17 exact constraints~\cite{Sun2015}, reducing but not eliminating self-interaction error in comparison to LDA and PBE. SCAN has been tested extensively on $d$- and $f$-electron materials, as reviewed in Ref.~\cite{zhang2025advances}. Without a Hubbard $U$, it reproduces the antiferromagnetic insulating ground state of La$_2$CuO$_4$, where LDA and PBE give a nonmagnetic metal~\cite{Lane2018}. 

Here, we find that magnetism dramatically enhances EPC: at 25\% hole doping, the total coupling strength reaches $\lambda\sim0.5$, nearly five times that of the nonmagnetic reference at the same doping. In turn, this enhancement produces pronounced quasiparticle kinks and characteristic peak--dip--hump (PDH) features in the computed spectral functions, with self-energy magnitudes comparable to experiment~\cite{Hongtao_PNAS}. We identify oxygen rotational phonons as the dominant contributors to the coupling in the 50--65~meV window, where they primarily generate the high-energy PDH features, while the commonly implicated oxygen-breathing modes contribute only weakly. Furthermore, under hole doping, the coupling strength increases to $\lambda\sim1.6$ at 37.5\%, where low-frequency phonons below 45~meV soften and couple strongly near the Fermi-surface nesting wave vector. At 56.25\%, where the Cu magnetic moments vanish and the circular polarization of the oxygen modes nearly disappears, $\lambda$ decreases to $\sim0.4$. Finally, we show that the rotational modes carry staggered angular momentum with no net angular momentum in the unit cell; however, a small time-reversal-breaking gyrotropic coupling on the Cu sites yields a finite phonon thermal Hall conductivity. Our study thus identifies which phonon branches couple to electrons in the magnetic state and how this coupling changes with doping, predictions that can be tested by momentum-resolved phonon measurements.

\section{Results }

We begin by calculating the EPC in pristine CaCuO$_2$ (CCO) using SCAN~\footnote{Our previous work has shown that the advanced SCAN functional yields more reliable electronic structures and phonon dispersions for cuprates than the conventional LDA and PBE functionals~\cite{Furness2018,Lane2018,ZhangYPANS,NingPRBYBCO6,NingJCPDFAYBCO6}. Our recent work has demonstrated the strong transferability of the \rrscan/SCAN density functional for predicting EPC both accurately and efficiently across a wide range of materials—from itinerant $s$/$p$-electron systems to strongly correlated $d$-electron compounds~\cite{wang_epc_rrscan,Zhang_LNO_EPC}} in two configurations: the $G$-type antiferromagnetic ($G$-AFM) ground state and a nonmagnetic (NM) reference configuration, which lies 73.5~meV/Cu higher in energy (SM). In the NM state, we obtain weak EPC, $\lambda \sim 0.1$, whereas the $G$-AFM state exhibits markedly stronger coupling. The weak EPC in the NM state is consistent with earlier PBE calculations~\cite{giustino2008cuprate}, while the strong EPC in the $G$-AFM state is consistent with a recent PBE+$U$ study~\cite{DFTU_EPC_LCO}. These results show that magnetic order substantially increases the EPC in CCO, and the $G$-AFM state therefore serves as the reference for the doped calculations below. A detailed analysis of the undoped case is given in the Supplementary Material (SM).

{\subsection{Evolution of electronic structure and magnetism with hole doping}
\label{sec:doping_electronic_structure}

\begin{figure*}[ht]\centering
\includegraphics[width=0.9\textwidth]{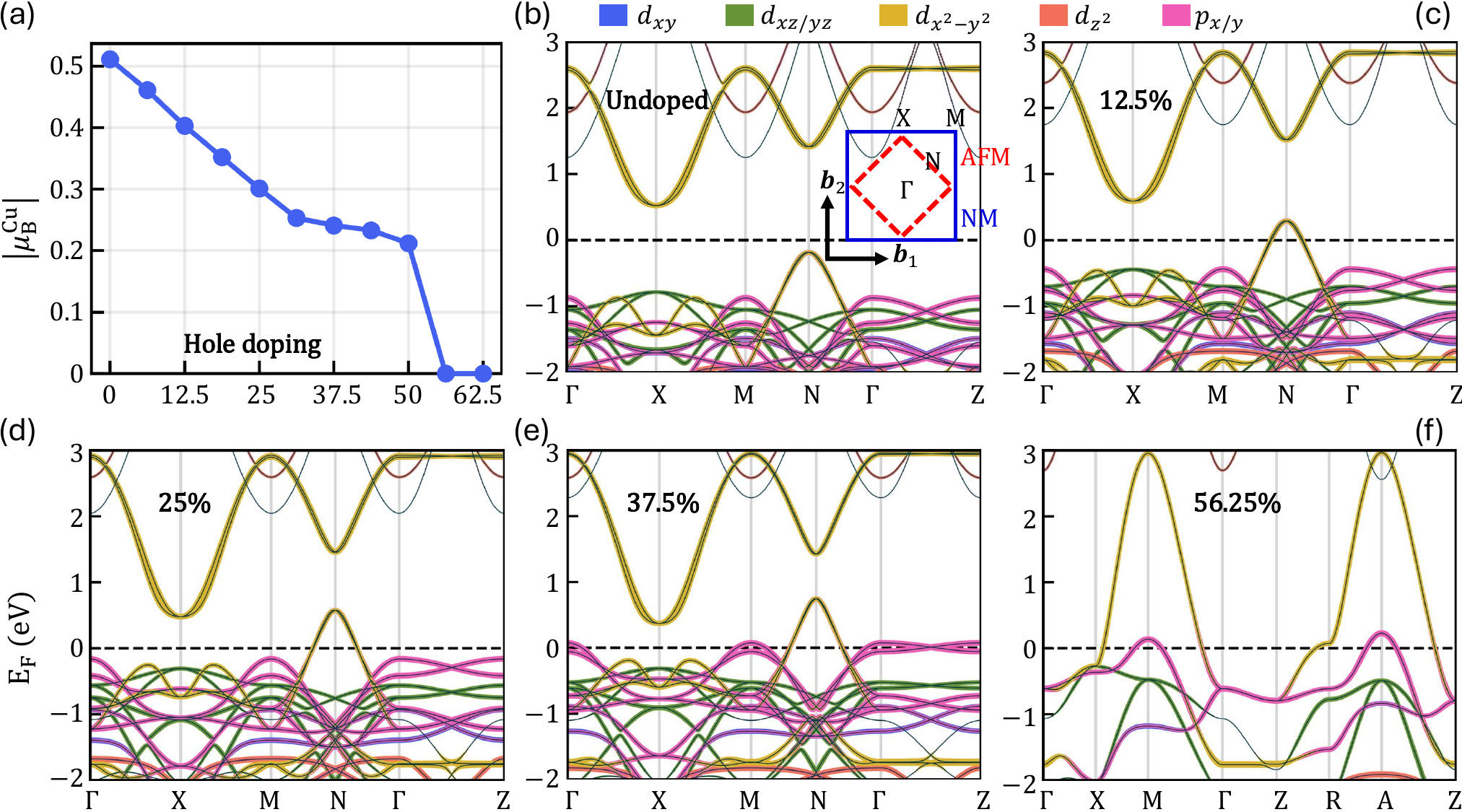}
\caption{\textbf{Cu magnetic moment and orbital-projected band structures of CaCuO$_2$ as functions of hole doping.}
(a) Mean absolute Cu magnetic moment, $\bar{m}_{\mathrm{Cu}}$, versus hole doping, calculated in steps of 6.25\%; lines connecting the calculated points are guides to the eye.
(b)--(e) Band structures of the $G$-AFM configuration at hole dopings of 0\%, 12.5\%, 25\%, and 37.5\%, respectively.
The inset in (b) shows the $k_z=0$ sections of the NM (solid blue) and $G$-AFM (dashed red) Brillouin zones (BZ); $\mathbf{b}_1$ and $\mathbf{b}_2$ are the in-plane reciprocal-lattice vectors of the NM cell, and $X=(\pi/a,0)$, $M=(\pi/a,\pi/a)$, and $N=(\pi/2a,\pi/2a)$, where $a$ is the in-plane NM lattice constant.
(f) Band structure of the NM configuration at 56.25\% hole doping; $Z$, $R$, and $A$ are high-symmetry points on the $k_z=\pi/c$ plane.
The Fermi level $E_{\mathrm F}$ is set to zero and indicated by the black dashed line. Colors denote projections onto the Cu $d_{xy}$ (blue), $d_{xz}/d_{yz}$ (green), $d_{x^2-y^2}$ (yellow), and $d_{z^2}$ (red) orbitals and the O $p_x/p_y$ (magenta) orbitals.}
\label{fig:cco_doping_electronic}
\end{figure*}

\begin{figure*}[ht]\centering
\includegraphics[width=\textwidth]{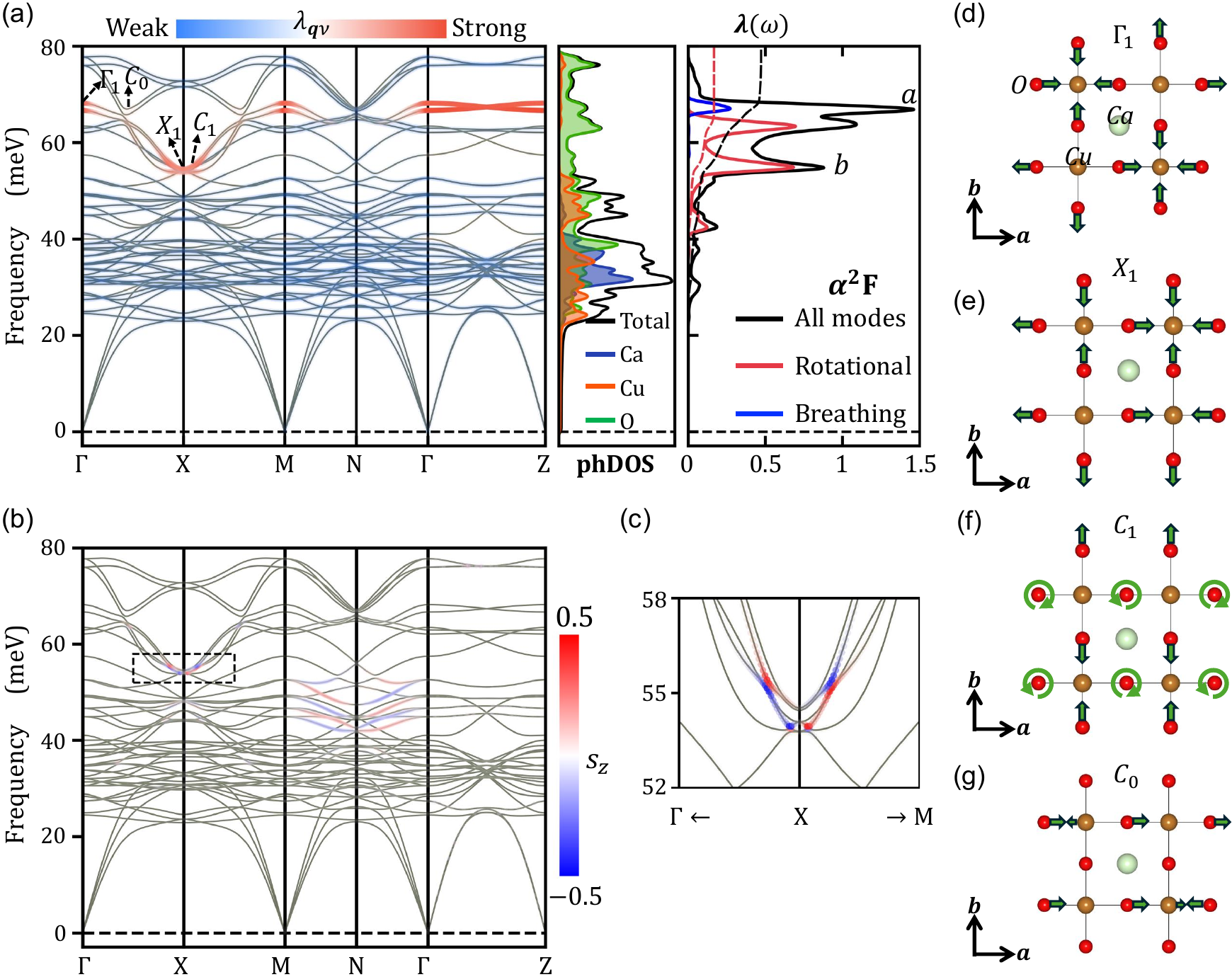}  %
\caption{\textbf{Calculated phonon dispersion and EPC of $G$-AFM configuration of CaCuO$_2$ with 25\% hole doping.}
(a) Left: folded phonon dispersion along high-symmetry lines of the NM BZ; the marker color represents the mode-resolved EPC strength $\lambda_{\mathbf{q}\nu}$. Arrows mark the modes $\Gamma_1$, $C_0$, $X_1$, and $C_1$ shown in (d)--(g), respectively. Middle: total (black) and atom-projected (Ca, blue; Cu, orange; O, green) phonon density of states (PhDOS). Right: Eliashberg spectral function $\alpha^2F(\omega)$ (solid lines) and cumulative EPC strength $\lambda(\omega)$ (dashed lines) for all modes (black) and for the rotational modes (red); the solid blue line is $\alpha^2F(\omega)$ of the breathing modes. A mode is classified as rotational if $\sum_{i\in\mathrm{O}}|\mathbf{e}^{\dagger}_{i,\mathbf{q},\mu} S^{z}\mathbf{e}_{i,\mathbf{q},\mu}| \geq 0.1$, where the sum runs over all eight O sites in the unit cell [see Eq.~(\ref{eq:phonon_circular_polarization})]. Peaks $a$ and $b$ are discussed in the text. A $60\times60\times60$ $k$-mesh and a $30\times30\times30$ $q$-mesh were used to sample the full BZ. (b) Phonon dispersion color-coded by the $z$-component of the circular polarization, summed over four representative O sites [O$_1$, O$_3$, O$_6$, and O$_8$; numbered as in Fig.~S3(c) of the SM]. Red, blue, and gray denote right-handed ($s^{z}_{\mathbf{q},\mu}>0$), left-handed ($s^{z}_{\mathbf{q},\mu}<0$), and non-polarized ($s^{z}_{\mathbf{q},\mu}=0$) modes, respectively. (c) Close-up of the region marked by the dashed rectangle in (b), near $X$. (d)--(g) Top views ($ab$ plane) of the $\Gamma_1$, $X_1$, $C_1$, and $C_0$ modes marked in (a); green arrows indicate the atomic displacements. In (f), neighboring O atoms rotate alternately clockwise and counterclockwise about their equilibrium positions. Brown, light green, and red balls represent Cu, Ca, and O atoms, respectively.}
\label{fig:cco_doping}
\end{figure*}

Before examining the doping dependence of EPC, we calculate the Cu magnetic moments and band structures of CCO for hole doping from 0 to 62.5\% in steps of 6.25\%. Hole doping is simulated by assigning a non-integer nuclear charge to the Ca site~\cite{NNC_1D}, and the crystal structure is fully relaxed at each doping level. Figure~\ref{fig:cco_doping_electronic}(a) shows the mean absolute Cu magnetic moment, $\bar{m}_{\mathrm{Cu}}$, as a function of hole doping. In the undoped compound, $\bar{m}_{\mathrm{Cu}}=0.511\,\mu_{\mathrm{B}}$, close to the ordered moment of $0.51\pm0.05\,\mu_{\mathrm{B}}$ measured by neutron diffraction on the infinite-layer compound Ca$_{0.85}$Sr$_{0.15}$CuO$_2$~\cite{Vaknin1989}. Up to 31.25\% doping, $\bar{m}_{\mathrm{Cu}}$ decreases almost linearly, to $0.403\,\mu_{\mathrm{B}}$, $0.301\,\mu_{\mathrm{B}}$, and $0.253\,\mu_{\mathrm{B}}$ at 12.5\%, 25\%, and 31.25\%, respectively. Between 31.25\% and 50\%, it decreases by only $0.041\,\mu_{\mathrm{B}}$, to $0.241\,\mu_{\mathrm{B}}$ at 37.5\% and $0.212\,\mu_{\mathrm{B}}$ at 50\%. This slower decrease is consistent with holes being doped into both the Cu $3d_{x^2-y^2}$ and the planar O $p_x/p_y$ states in this range [see Fig.~\ref{fig:cco_doping_electronic}(e)], so that fewer holes enter the Cu $3d_{x^2-y^2}$ states. At 56.25\% and 62.5\%, calculations initialized in the $G$-AFM configuration converge to nonmagnetic solutions.

In our calculations, the Cu magnetic moments remain finite up to 50\% doping. In fact, measurements of magnetic excitations likewise indicate that antiferromagnetic correlations in hole-doped cuprates persist well beyond optimal doping. Inelastic neutron scattering on overdoped La$_{2-x}$Sr$_x$CuO$_4$ detects antiferromagnetic spin excitations at $x=0.25$, with a spectral weight that decreases between $x=0.25$ and $0.30$~\cite{Wakimoto2007Overdoped}, and resonant inelastic x-ray scattering (RIXS) detects paramagnon excitations up to $x=0.40$~\cite{Dean2013Persistence}.
Calculations for the three-orbital Emery model also find that the instantaneous Cu magnetic moment decreases with hole doping for several charge-transfer energies~\cite{Peng2025EmeryMagnetism}.
Neutron scattering and RIXS measure dynamical spin correlations, and the Emery-model calculations give the instantaneous local moment, whereas the present calculations give the static Cu spin polarization of a $G$-AFM reference state. As discussed in our earlier study of YBa$_2$Cu$_3$O$_7$~\cite{ZhangYPANS}, the intertwined orders in cuprates appear in our SCAN calculations as a series of nearly degenerate AFM configurations with various stripe patterns; we use the $G$-AFM configuration as an approximation to this manifold of states.

Figures~\ref{fig:cco_doping_electronic}(b)--(f) show the orbital-resolved band structures for selected doping levels. The undoped compound is an insulator [Fig.~\ref{fig:cco_doping_electronic}(b)], with an indirect gap of approximately $0.57$~eV between the valence-band maximum (VBM) at $N$ and the conduction-band minimum (CBM) at $X$. Both the VBM and the CBM are dominated by Cu $3d_{x^2-y^2}$ states. At 12.5\%, 25\%, and 37.5\% hole doping, the lower $3d_{x^2-y^2}$-dominated band crosses the Fermi level near $N$. The calculations therefore capture the insulator-to-metal transition upon hole doping, as found previously with SCAN calculations for La$_{2-x}$Sr$_x$CuO$_4$~\cite{Furness2018}. With increasing doping, the energy separation between the lower and upper Hubbard $3d_{x^2-y^2}$ bands decreases, following the reduction of $\bar{m}_{\mathrm{Cu}}$. At 37.5\%, an O $p_x/p_y$-dominated band, which is nearly flat along $\Gamma$--$Z$, also crosses the Fermi level near $\Gamma$; the holes are thus shared between the Cu and planar O states. In the NM configuration at 56.25\% [Fig.~\ref{fig:cco_doping_electronic}(f)], plotted in the NM Brillouin zone (BZ), two bands cross the Fermi level: one dominated by Cu $3d_{x^2-y^2}$ and the other by O $p_x/p_y$. In addition, the Cu $3d_{x^2-y^2}$-dominated band has a saddle point at $R$, approximately $0.07$~eV above the Fermi level, which produces a van Hove singularity in the density of states near the Fermi level. Because $\lambda\propto N(E_F)$~\cite{McMillan1968}, such a singularity could enhance the EPC near this doping.

\par}

\subsection{EPC at 25\% hole doping}
\label{sec:epc_25}

\subsubsection{Phonon modes with strong electron--phonon coupling}

{Here we identify the phonon modes that couple most strongly to electrons at 25\% doping. The results for 37.5\% and 56.25\% are presented in Secs.~\ref{sec:epc_375} and \ref{sec:epc_5625}, the EPC strengths and estimated $T_{\mathrm c}$ values are compared across dopings in Sec.~\ref{sec:doping_tc}, and the results for the undoped compound and for 12.5\% doping are given in the SM (Secs.~S-III and S-IV).
\par}

The calculated phonon dispersion, phonon density of states (PhDOS), and isotropic Eliashberg spectral function $\alpha^2F(\omega)$ are shown in the left, middle, and right panels of Fig.~\ref{fig:cco_doping}(a), respectively. The color scale on the phonon dispersions represents the mode-resolved EPC strength $\lambda_{q\nu}$. Several branches exhibiting strong coupling can be seen. Most prominent are the high-frequency oxygen stretching and circular modes, which dominate $\alpha^2F(\omega)$ in the 50--70 meV range. Lower-frequency modes involving Cu and Ca motions also contribute, giving rise to weaker peaks near 42, 38, and 30 meV in $\alpha^2F(\omega)$ [see Fig.~\ref{fig:cco_doping}(a)]. Integrating over all phonons yields the total coupling strength $\lambda$ of $\sim$0.5—approximately five times larger than that in the NM configuration—highlighting a substantial enhancement of EPC via magnetic effects. The estimated $\hbar\omega_{\log}$ is 57.0~meV, and the McMillan--Allen--Dynes expression [Eq.~(\ref{eq:tc_doping})] gives $T_{\mathrm c}$ of 6--14~K for $\mu^*$ between 0.10 and 0.04. These results make it clear that the AFM state hosts a rich variety of phonon modes in cuprates, and that there are substantial differences in EPC strength between the NM and AFM phases.

Representative phonon eigenmodes that contribute strongly to the EPC in CCO are shown in Fig.~\ref{fig:cco_doping}(d)--(g). Panels (d)--(f) sample points on the O-derived branches between 50–70 meV [see labeled points in the left panel of Fig.~\ref{fig:cco_doping}(a)]. At high-symmetry points, the $\Gamma_{1}$ and $X_{1}$ modes correspond to an in-plane full-breathing and a quadrupolar oxygen vibration motion, respectively. Away from the high-symmetry lines, the phonon eigenvectors describe in-plane rotational motions of the ions within the $ab$-plane. For example, the mode in panel (f) combines rotational phonons along $z$ with a half-breathing mode along $y$. We also find a characteristic softening associated with the $C_{0}$ branch at $\mathbf{q}=(0.22,\,0,\,0)$ around $\sim 65$~meV, close to the experimentally reported charge-density-wave (CDW) wave vector $\mathbf{q}_{\mathrm{CDW}}=(0.25,\,0,\,0)$. The $C_{0}$ mode is dominated by a half-breathing character involving O and Cu motions [Fig.~\ref{fig:cco_doping}(g)]. Connection to the experimental CDW phenomenology is further discussed in the SM.

\begin{figure}[ht]\centering
\includegraphics[width=.9\columnwidth]{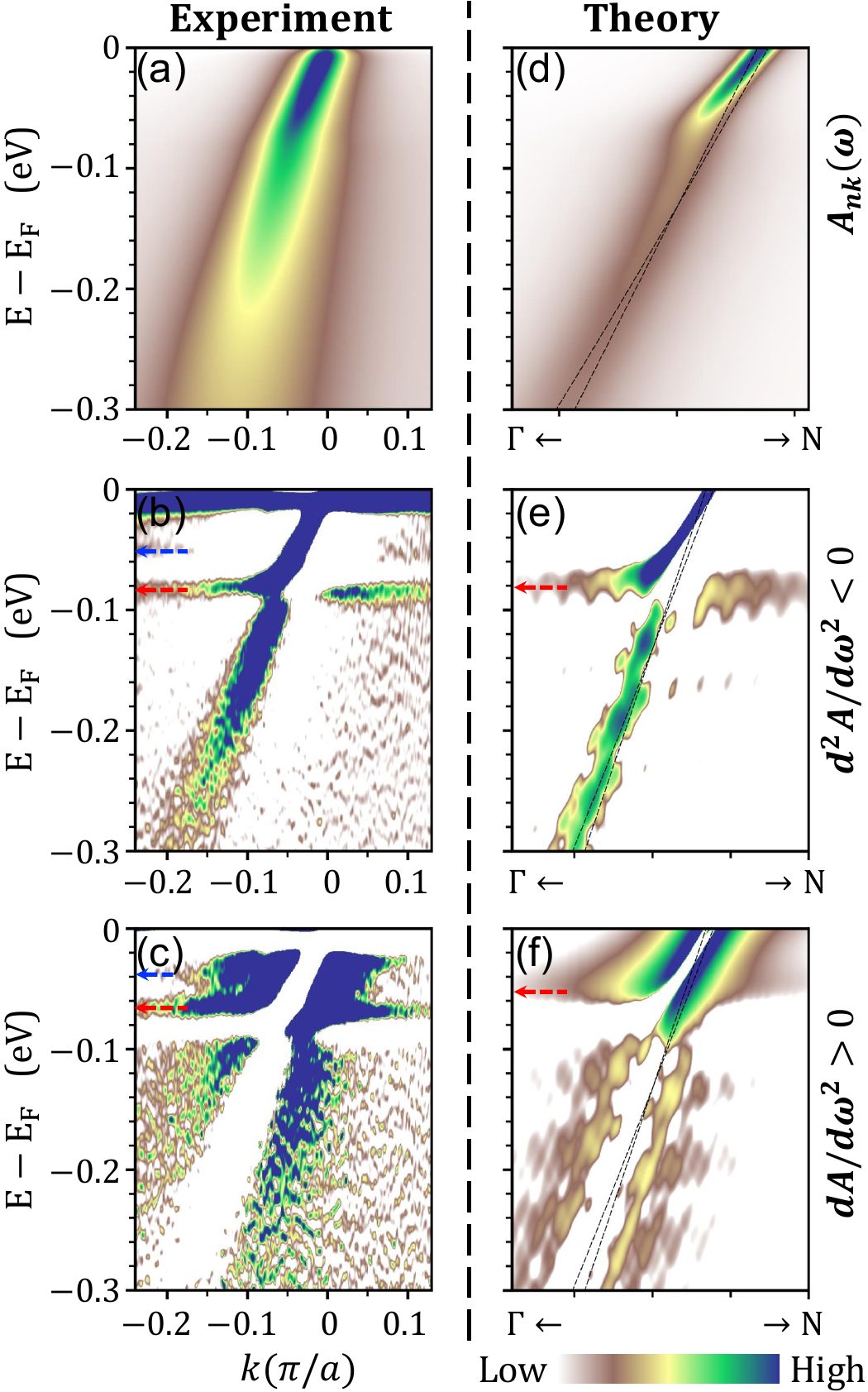}  %
\caption{\textbf{Comparison of the experimental and calculated spectral functions}. (a)--(c) Measured ARPES spectra of Bi2201 at a hole doping of approximately 20\% and $T=20$~K, taken from Ref.~\cite{Hongtao_PNAS}: (a) spectral function, (b) magnitude of the negative part and (c) positive part of the second derivative with respect to energy. Red (blue) arrows mark the high- (low-)energy humps and dips. In (a)--(c), $k=0$ is the Fermi momentum, where the measured band crosses the Fermi level. (d)--(f) Same as (a)--(c), calculated along $\Gamma$--$N$ for CCO at 25\% doping with all phonon modes; red arrows mark the high-energy hump and dip. (g)--(i) Same as (d)--(f), with only the rotational phonon modes included.}
\label{fig:cco_kink}
\end{figure}

We quantify the circular polarization (angular momentum) of each phonon mode following Refs.~\cite{Qian2015/PhysRevLett.115.115502,nanolett2025/chiral}:
\begin{equation}
s^{\alpha}_{\mathbf{q},\mu}
=\sum_{i=1}^{N}\sum_{\beta,\gamma} e^{*}_{i\beta,\mathbf{q}\mu}\,(S^{\alpha})_{\beta\gamma}\,e_{i\gamma,\mathbf{q}\mu},
\label{eq:phonon_circular_polarization}
\end{equation}
where $e_{i\beta,\mathbf{q}\mu}$ is the Cartesian component $\beta$ ($=x,y,z$) of atom $i$ in the eigenvector of branch $\mu$ at wave vector $\mathbf{q}$, and $(S^{\alpha})_{\beta\gamma}=-i\epsilon_{\alpha\beta\gamma}$ are the spin-1 matrices. The sign of $s^{\alpha}_{\mathbf{q},\mu}$ gives the handedness of the mode with respect to the $\alpha$ axis: $s^{\alpha}_{\mathbf{q},\mu}>0$ ($<0$) corresponds to a right-handed (left-handed) circularly polarized phonon.  

Figure~\ref{fig:cco_doping}(b) displays the $z$ component of the circular polarization of the oxygen-dominated modes. We identify three sets of rotational branches: two oxygen optical branches, one near 55~meV along $\Gamma$--$X$--$M$ and another near 45~meV along $M$--$N$--$\Gamma$ [BZ in Fig.~\ref{fig:cco_doping_electronic}(b)], and a low-energy acoustic branch involving Ca motions, similar to that of the doped NM phase (Fig.~S5). Most phonon branches carry zero net circular polarization, with equal numbers of O atoms rotating clockwise (CW) and counterclockwise (CCW) [Fig.~\ref{fig:cco_doping}(f)]. Figures~\ref{fig:cco_doping}(b) and (c) show the net circular polarization, defined as the sum of the circular polarizations of four representative O atoms [O$_1$, O$_3$, O$_6$, and O$_8$; numbered as in Fig.~S3(c) of the SM] of the eight in the AFM unit cell; the site-resolved circular polarization (Fig.~S6) shows the CW and CCW rotations of the individual atoms. 

The particular rotational oxygen mode in Fig.~\ref{fig:cco_doping}(f) can be described by a single angle that measures the rotation of oxygen atoms about the local $c$ axis, with alternating signs on neighboring sites. %
For small angles, the energy is that of a harmonic rotor, which is qualitatively consistent with the calculated eigenvectors [Fig.~\ref{fig:cco_doping}(f)]. Similar descriptions have been used for quartz~\cite{Ueda2023} and (Mn,Ni)$_3$TeO$_6$~\cite{Okamoto2025}. Because the rotation angle changes the local Cu--O geometry, it can modulate the hopping, which provides a plausible qualitative explanation for the enhanced coupling of these modes. The quantitative phonon properties and coupling strengths are determined by the full first-principles calculations.

We emphasize that the rotational modes with strong EPC emerge only in the doped AFM phase. In the NM phase, many optical branches are degenerate and involve linear atomic motions, but broken time-reversal symmetry in the AFM state lifts these degeneracies and selects circularly polarized eigenstates, making the circular polarization directly manifest in the phonon modes. Note also that AFM and NM phases both host rotational character in the two degenerate transverse acoustic modes along $\Gamma-Z$, arising from vibrations of Ca atoms (Figs.~S5 and S6); these acoustic rotational branches, however, couple only weakly to the electrons.

A comparison of panels (a) and (b) of Fig.~\ref{fig:cco_doping} shows that rotational phonons are the modes most strongly coupled to electrons. The mode-decomposed $\alpha^2F(\omega)$ in the right panel of Fig.~\ref{fig:cco_doping}(a) reveals that rotational oxygen vibrations contribute about one third of the total EPC strength ($\lambda_{\mathrm{rot}}\approx0.17$) and dominate peak $b$ of $\alpha^2F(\omega)$, whereas peak $a$ at 67~meV is contributed by non-rotational modes. The remaining coupling comes from non-rotational modes; among these, the full- and half-breathing oxygen modes [blue line in the right panel of Fig.~\ref{fig:cco_doping}(a)] contribute only $\lambda_{\mathrm{breathing}}\approx0.01$. These results challenge the widely held view that breathing modes play a dominant role in cuprates.

\subsubsection{Low-energy kinks}

Figures~\ref{fig:cco_kink}(d)--(f) show the calculated spectral functions along the nodal direction of hole-doped CCO. The pronounced band renormalization and quasiparticle broadening evident in the 40--80~meV energy range [Fig.~\ref{fig:cco_kink}(d)] are consistent with the ARPES nodal dispersion kinks in Bi2201 [Fig.~\ref{fig:cco_kink}(a)]~\cite{Hongtao_PNAS}. A second-derivative analysis highlights a hump feature near  80~meV [Fig.~\ref{fig:cco_kink}(e)] and a dip feature near 50~meV [Fig.~\ref{fig:cco_kink}(f)], close to the experimental values of 86 and 67~meV, respectively [Figs.~\ref{fig:cco_kink}(b) and (c)]~\cite{Hongtao_PNAS}. These results demonstrate that our calculations correctly capture the well-known peak--dip--hump features in cuprates. 

Our calculations show a single hump and a single dip in the spectra [Figs.~\ref{fig:cco_kink}(e) and (f)], while experiments on Bi-based cuprates report two humps and two dips [Figs.~\ref{fig:cco_kink}(b) and (c)], although the second hump and dip are relatively weak and were not observed in earlier, lower-resolution ARPES experiments. This discrepancy likely reflects the material difference between infinite-layer CCO and the single-layer cuprate Bi2201 studied experimentally. Note also that multiple competing orders present in doped cuprates~\cite{ZhangYPANS,Paper1} (e.g., stripe and charge-density-wave (CDW) phases) are not included in our single-phase calculations and may contribute to the observed differences.  

\begin{figure}[ht]\centering
	\includegraphics[width=\columnwidth]{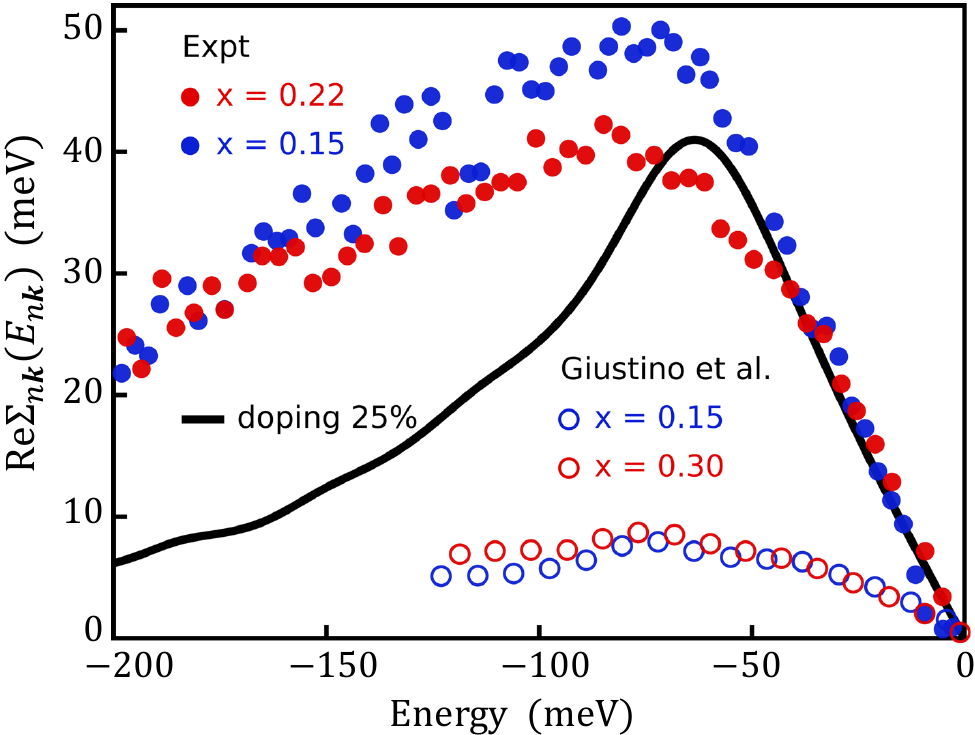}
	\caption{\textbf{Real part of the electron self-energy along the nodal direction.}
		Black line: Calculated $\mathrm{Re}\,\Sigma_{n\mathbf{k}}(E_{n\mathbf{k}})$ from EPC in $G$-AFM CaCuO$_2$ at 25\% hole doping.
		Filled circles: experimental values for La$_{2-x}$Sr$_x$CuO$_4$ at $x=0.15$ (blue) and $0.22$ (red) at 20~K, extracted in Ref.~\cite{giustino2008cuprate} from the ARPES data of Ref.~\cite{ShenZX_Nature_70}.
		Open circles: $\mathrm{Re}\,\Sigma_{n\mathbf{k}}(E_{n\mathbf{k}})$ calculated from EPC for La$_{2-x}$Sr$_x$CuO$_4$ at $x=0.15$ (blue) and $0.30$ (red) in a NM configuration using the PBE functional~\cite{giustino2008cuprate}. Energies are measured from the Fermi level.}
	\label{fig:cco_selfenergy}
\end{figure}

\begin{figure*}[ht]\centering
\includegraphics[width=\textwidth]{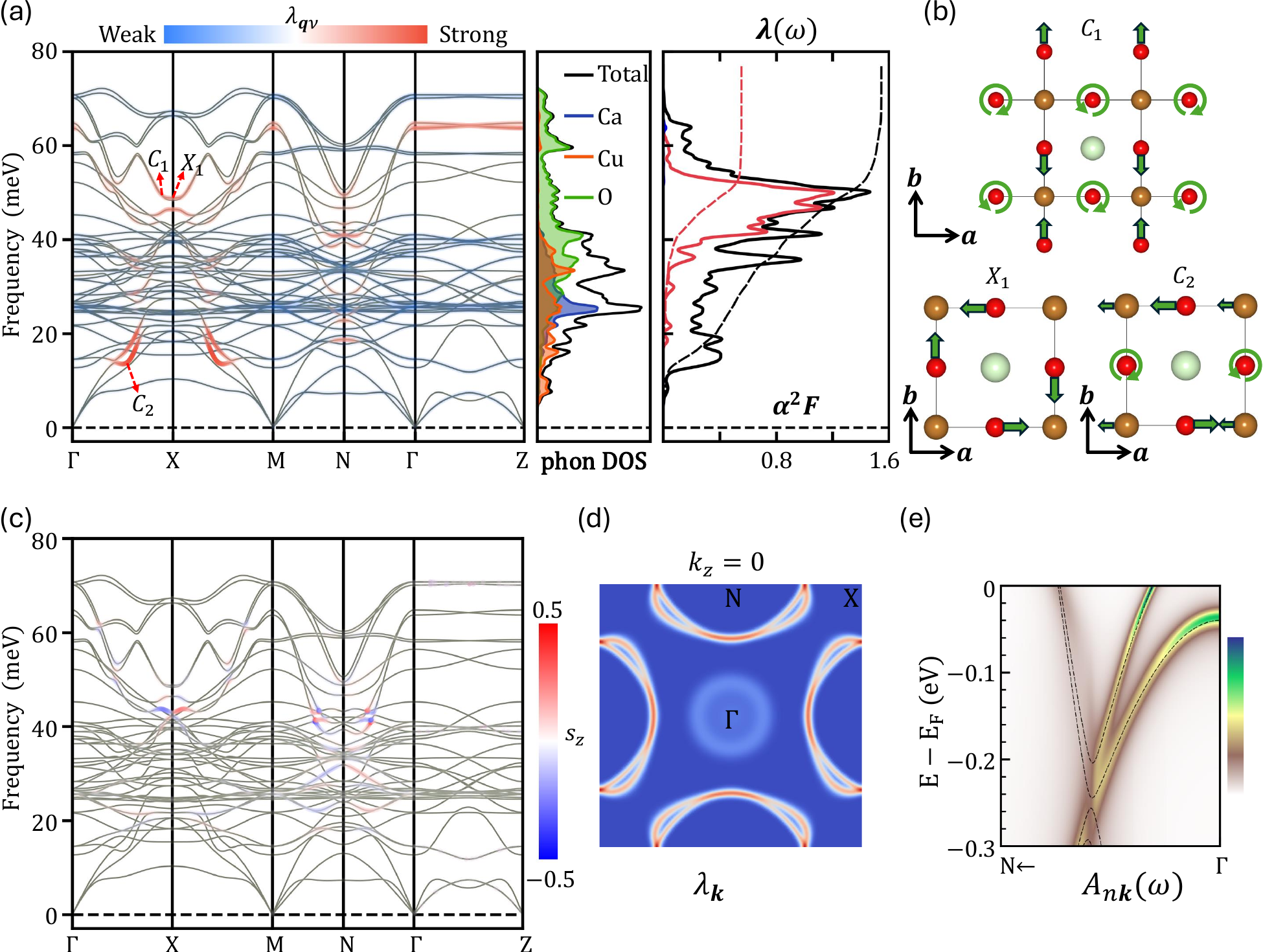}
\caption{\textbf{Phonons and electron--phonon coupling in $G$-AFM CaCuO$_2$ at 37.5\% hole doping.}
(a) Left: phonon dispersion along high-symmetry lines in the NM BZ; the marker color represents the mode-resolved EPC strength $\lambda_{\mathbf{q}\nu}$. Middle: total (black) and atom-projected (Ca, blue; Cu, orange; O, green) PhDOS. Right: Eliashberg spectral function $\alpha^2F(\omega)$ (solid lines) and cumulative EPC strength $\lambda(\omega)$ (dashed lines) for all modes (black) and for the rotational oxygen modes (red); the solid blue line is $\alpha^2F(\omega)$ of the breathing modes.
(b) Top views ($ab$ plane) of the $C_1$, $X_1$, and $C_2$ modes; straight arrows indicate linear displacements, and circular arrows indicate the sense of rotation of the O atoms.
(c) Phonon dispersion color-coded by the $z$ component of the oxygen circular polarization [Eq.~(\ref{eq:phonon_circular_polarization})].
(d) Momentum-resolved EPC strength $\lambda_{\mathbf k}$, summed over bands, on the Fermi surface in the $k_z=0$ plane of the $G$-AFM BZ.
(e) Electron spectral function $A_{n\mathbf k}(\omega)$ along $\Gamma$--$N$; dashed lines are the bare bands.}
\label{fig:cco_epc_375}
\end{figure*}

{\subsubsection{Real part of the electron self-energy}
\label{sec:selfenergy_25}

The spectral functions in Fig.~\ref{fig:cco_kink} identify the energies of the kink features. To quantify the strength of the phonon-induced renormalization, Fig.~\ref{fig:cco_selfenergy} shows the real part of the calculated electron self-energy along the nodal direction at 25\% doping, $\mathrm{Re}\,\Sigma_{n\mathbf{k}}(E_{n\mathbf{k}})$, together with experimental values for La$_{2-x}$Sr$_x$CuO$_4$ at $x=0.15$ and $0.22$. Giustino \textit{et al.}~\cite{giustino2008cuprate} extracted these values from the ARPES dispersions of Ref.~\cite{ShenZX_Nature_70} and compared them with their generalized-gradient-approximation (GGA) results for $x=0.15$ and $0.30$, obtained in a nonmagnetic configuration~\cite{giustino2008cuprate}; both data sets are taken from Ref.~\cite{giustino2008cuprate}.

Between the Fermi level and $-50$~meV, the calculated electronic self-energy closely follows the experimental values. At $-25$~meV, the calculated value is approximately 17~meV, compared with approximately 18~meV in experiment; at $-50$~meV, it is approximately 36~meV, compared with 40 and 31~meV for $x=0.15$ and $0.22$, respectively. The calculated $\mathrm{Re}\,\Sigma$ reaches a maximum of 41~meV at $-64$~meV, whereas the experimental maxima are approximately 50~meV near $-80$~meV for $x=0.15$ and 42~meV near $-85$~meV for $x=0.22$. The calculated maximum is thus close to the overdoped value, about 20\% smaller than the optimally doped value, and 17--21~meV closer to the Fermi level.

In contrast, the earlier GGA self-energies of Ref.~\cite{giustino2008cuprate}, obtained in a NM configuration, reach only 8--9~meV, about one-fifth of the experimental maxima. Unlike Ref.~\cite{giustino2008cuprate}, the present calculations use an $G$-AFM configuration with the SCAN functional. By including GW corrections to the EPC matrix elements, recent GWPT calculations for La$_{2-x}$Sr$_x$CuO$_4$ at $x=0.15$ and $0.30$, also in a nonmagnetic configuration, similarly obtain a three times larger phonon-induced self-energy than DFT~\cite{ZhengluGWPTCuprate}. However, in the GWPT calculations~\cite{ZhengluGWPTCuprate}, this peak arises from the Cu--O half- and full-breathing modes, whereas the present maximum, 41~meV at $-64$~meV, lies at a lower binding energy, where rotational oxygen modes dominate the coupling.

At larger binding energies, the calculated self-energy decreases more rapidly than the measured one from experiments: at $-150$~meV, it is approximately 12~meV, whereas the experimental values are 30--35~meV. Part of this difference may come from the experimental values being obtained relative to an assumed linear bare band dispersion~\cite{giustino2008cuprate,ZhengluGWPTCuprate}. The remaining difference suggests that electronic correlations and spin fluctuations~\cite{Byczuk2007,Dahm2009} should be important for understanding the electronic self-energy at high energy. The comparison is also limited by the different doping levels and by the absence of apical oxygen in CaCuO$_2$, unlike in La$_{2-x}$Sr$_x$CuO$_4$.

\par}
{\subsection{EPC at 37.5\% hole doping: phonon softening and low-frequency coupling}
\label{sec:epc_375}

The calculated phonon dispersion, PhDOS, and $\alpha^2F(\omega)$ at 37.5\% doping are shown in the left, middle, and right panels of Fig.~\ref{fig:cco_epc_375}(a), respectively. Compared with 25\% doping [Fig.~\ref{fig:cco_doping}(a)], the entire phonon spectrum is softer: at $\Gamma$, the highest optical branch decreases from $\sim$78 to 71~meV, the lowest optical branch from $\sim$25 to 13~meV, and the $X_1$ mode from $\sim$54 to 47~meV. The electron--phonon coupling shifts accordingly to lower-energy branches. The main peak of $\alpha^2F(\omega)$ lies near 50~meV, with additional peaks between 30 and 45~meV and appreciable coupling between 10 and 20~meV. Phonons below 20~meV contribute $\lambda\approx0.38$, whereas their contribution at 25\% hole doping is negligible. In contrast, phonons above 60~meV, which dominate the EPC at 25\%, contribute only $\lambda\approx0.03$. Integrating over all phonons yields $\lambda\approx1.57$, about three times the value at 25\%. The estimated $\omega_{\log}$ is 31.2~meV, and the predicted $T_{\mathrm c}$ is 43--50~K for $\mu^*$ between 0.10 and 0.04 (Table~\ref{tab:doping_tc}). The spectral function along $\Gamma$--$N$ [Fig.~\ref{fig:cco_epc_375}(e)] shows the band renormalization due to EPC near $E_F$: the imaginary part of the electron self-energy strongly broadens both the O $p_x/p_y$ and the $d_{x^2-y^2}$ bands near $\Gamma$, while the real part produces a kink in the Cu $3d_{x^2-y^2}$ band at binding energies of 50--80~meV.

Representative modes are shown in Fig.~\ref{fig:cco_epc_375}(b). The strongly coupled low-frequency modes lie between about 12 and 20~meV along $\Gamma$--$X$ and $X$--$M$. The mode $C_2$ at 12.3~meV combines linear Cu displacements with elliptical motions of some of the O atoms; its squared mass-weighted eigenvector is distributed over Cu (49\%), Ca (33\%), and O (18\%). In mode $C_1$ near $X$, the O atoms on one set of Cu--O bonds rotate in the $ab$ plane while those on the other set move along the bonds, and $X_1$ is a quadrupolar O bond-stretching mode, as at 25\% doping. Figure~\ref{fig:cco_epc_375}(c) displays the $z$ component of the oxygen circular polarization. The rotational branches are concentrated between about 40 and 50~meV near $X$ and along $M$--$N$--$\Gamma$. The mode-decomposed $\alpha^2F(\omega)$ [red curves in Fig.~\ref{fig:cco_epc_375}(a)] shows that rotational oxygen modes contribute between 38 and 53~meV and account for more than $\sim$70\% of $\alpha^2F$ at its 50~meV peak. 

To understand the origin of the enhanced $\lambda$ at 37.5\% hole doping, we write $\lambda=N(E_F)\langle I^2\rangle/(M\langle\omega^2\rangle)$~\cite{McMillan1968}, where $N(E_F)$ is the density of states at the Fermi level, $\langle I^2\rangle/M$ is the Fermi-surface average of the squared EPC matrix element divided by the ionic mass, and $\langle\omega^2\rangle$ is the mean squared phonon frequency weighted by $\alpha^2F(\omega)$. The enhancement comes mainly from phonons below 45~meV, whose contribution increases from 0.06 at 25\% to 1.09 at 37.5\%, whereas the contribution of phonons above 45~meV changes only from 0.41 to 0.48. Decomposing this enhancement for the phonons below 45~meV, we find a factor of 1.9 from $N(E_F)$, which increases from 1.17 to 2.19 states/eV per spin, a factor of 1.6 from the phonon softening, and a factor of 6 from $\langle I^2\rangle/M$. The enhanced $\lambda$ at 37.5\% therefore originates mainly from stronger EPC matrix elements of the low-frequency phonons, together with the larger $N(E_F)$ and the phonon softening. 

The larger $\langle I^2\rangle/M$ of the low-frequency phonons reflects the change of the Fermi surface at 37.5\% hole doping, as shown by the nesting functions in Figs.~S9(b) and S9(e) of the SM. At 25\%, the Fermi surface consists only of the Cu $3d_{x^2-y^2}$ sheets around $N$. Along $\Gamma$--$X$, its nesting function has two peaks, A at $\mathbf q=(0.25,0)$, which corresponds to $Q_{\mathrm{CDW}}$, and B near $X$, with maximum values of 4.43 and 6.34, respectively, and these sheets couple mainly to the in-plane rotational oxygen modes above 50~meV that modulate the Cu--O bonds (Sec.~\ref{sec:epc_25}). At 37.5\% doping, the O $p_x/p_y$ band crosses the Fermi level and forms a small pocket around $\Gamma$ [Fig.~\ref{fig:cco_epc_375}(d)], in addition to the Cu $3d_{x^2-y^2}$ sheets around $N$. Peak A moves to $\mathbf q=(0.31,0)$, and its maximum increases by a factor of 2.5, to 11.11, exceeding that at B, which increases to 9.53. The coupling of the low-frequency branches is concentrated near $\mathbf q\approx(0.28,0)$, midway along $\Gamma$--$X$, and is nearly independent of $q_z$. The mode $C_2$ lies near this wave vector and modulates the Cu--O bonds. The $\Gamma$ pocket, in contrast, produces strong nesting at small $q$ but almost no coupling, so the O band contributes mainly through $N(E_F)$. Consistently, the momentum-resolved coupling $\lambda_{\mathbf k}$ in Fig.~\ref{fig:cco_epc_375}(d) shows that the strong EPC comes mainly from the Cu $3d_{x^2-y^2}$ sheets rather than from the O $p_x/p_y$ pocket. The larger $\langle I^2\rangle/M$ at 37.5\% thus arises near the nesting wave vector A, where the nesting is 2.5 times stronger and the coupling of the low-frequency branches is two orders of magnitude stronger than at 25\%.

Relative to 25\% doping, the low-frequency branches soften markedly; for example, the $C_2$ mode softens from $\sim$23 to 12.3~meV near the nesting wave vector, where the coupling of these branches is strongest. This coincidence suggests that the softening of this branch is enhanced by the nesting-assisted EPC: the coupling lowers the phonon frequency, and the lower frequency in turn increases $\lambda_{\mathbf q\nu}$. The softening of the other branches, in contrast, is not specific to the nesting wave vector; it also occurs, for example, for the optical modes at $\Gamma$. This softening may be related to stronger screening by the additional states at the Fermi level~\cite{FelicianoRMP2017} and to the expansion of the $c$ axis from 6.56 to 7.01~\AA.  To complete the picture, we note that many acoustic and low-energy optical modes display rotational Ca angular momentum along $x$ or $y$ directions, but any relevance for cuprate physics is currently unknown.

These results show that the strong EPC at 37.5\% arises from low-frequency phonons that couple strongly at the nesting wave vector midway along $\Gamma$--$X$, together with the larger density of states at the Fermi level and an overall phonon softening. Strong coupling to low-frequency phonon modes was also found in infinite-layer nickelates, where it involves a flat Ni $3d_{z^2}$ band~\cite{Zhang_LNO_EPC}.
\par}

{\subsection{EPC at 56.25\% hole doping: oxygen breathing modes in the nonmagnetic state}
\label{sec:epc_5625}

\begin{figure*}[ht]\centering
\includegraphics[width=.8\textwidth]{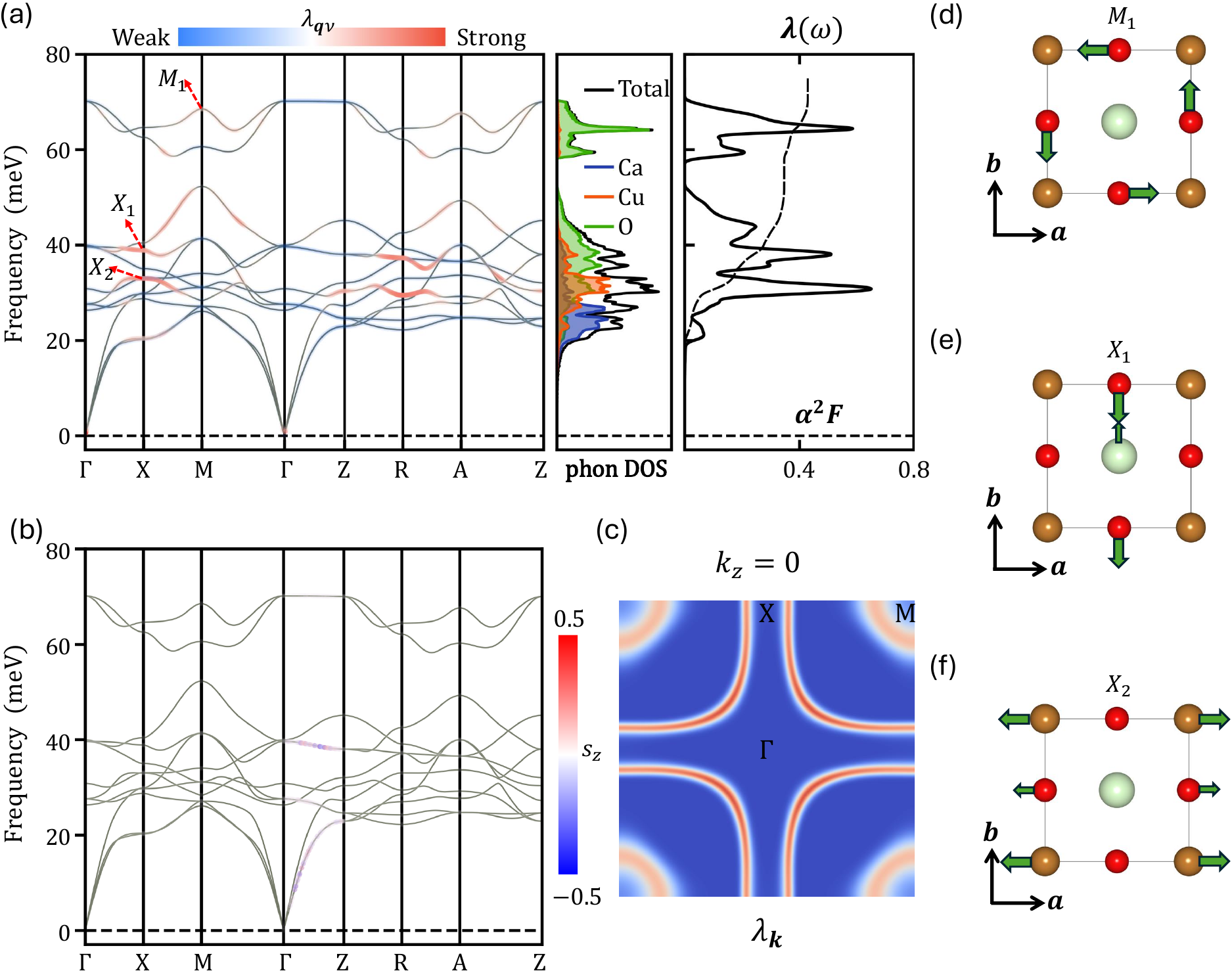}
\caption{\textbf{Phonons and electron--phonon coupling in nonmagnetic CaCuO$_2$ at 56.25\% hole doping.}
(a) Left: phonon dispersion along high-symmetry lines, with the marker color indicating the mode-resolved EPC strength $\lambda_{\mathbf q\nu}$. Middle: atom-projected phonon density of states (total, Ca, Cu, and O). Right: Eliashberg spectral function $\alpha^2F(\omega)$ (solid curve) and cumulative coupling $\lambda(\omega)$ (dashed curve).
(b) Phonon dispersion colored by the $z$ component of the oxygen circular polarization, Eq.~(\ref{eq:phonon_circular_polarization}).
(c) Momentum-resolved EPC strength $\lambda_{\mathbf k}$, summed over bands, on the Fermi surface in the $k_z=0$ plane.
(d)--(f) In-plane displacement patterns of the modes $M_1$, $X_1$, and $X_2$ marked in (a). Brown, green, and red balls represent Cu, Ca, and O atoms, respectively.}
\label{fig:cco_epc_5625}
\end{figure*}

At 56.25\% hole doping, the local magnetic moments of Cu vanish, and both the Cu $3d_{x^2-y^2}$- and O $p_x/p_y$-dominated bands cross $E_F$ [Fig.~\ref{fig:cco_doping_electronic}(f)]. The calculated phonon dispersion, PhDOS, and $\alpha^2F(\omega)$ are shown in Fig.~\ref{fig:cco_epc_5625}(a). Compared with 37.5\% doping, the strongly coupled low-frequency branches are absent: $\alpha^2F(\omega)$ vanishes below $\sim$20~meV, and the coupling comes mainly from optical branches between 30 and 70~meV. The two largest peaks of $\alpha^2F(\omega)$ lie near 31 and 64~meV, followed by weaker peaks near 38 and 45~meV. Integrating over all phonons yields $\lambda\approx0.42$, about one quarter of the value at 37.5\% and about four times the value of 0.1 in the undoped nonmagnetic phase (SM). The estimated $\omega_{\log}$ is 38.2~meV, and the predicted $T_{\mathrm c}$ is 2.5--7.0~K for $\mu^*$ between 0.10 and 0.04 (Table~\ref{tab:doping_tc}).

Representative modes are shown in Figs.~\ref{fig:cco_epc_5625}(d)--(f). The branches between 58 and 70~meV are almost entirely O-derived; their sharp PhDOS peak near 64~meV coincides with the 64~meV peak of $\alpha^2F(\omega)$. These branches couple most strongly at about 59~meV midway along $X$--$M$ and $R$--$A$; an example is the mode $M_1$ at $\sim$68~meV, a quadrupolar O bond-stretching mode. The two largest peaks of $\alpha^2F(\omega)$ come from branches between 30 and 40~meV, to which both Cu and O contribute. In the mode $X_2$ at $\sim$33~meV, the Cu atoms and the O atoms on the bonds along $b$ move along $a$, which modulates the Cu--O bond lengths along $a$. In the mode $X_1$ at $\sim$39~meV, the O atoms on the bonds along $a$ move perpendicular to the bonds, and the Ca atoms move along $b$ with a smaller amplitude.

Figure~\ref{fig:cco_epc_5625}(b) shows that the oxygen circular polarization nearly vanishes on all branches except for weak signals along $\Gamma$--$Z$, in contrast to the rotational branches at 25\% and 37.5\% doping. The momentum-resolved coupling $\lambda_{\mathbf k}$ [Fig.~\ref{fig:cco_epc_5625}(c)] is nearly uniform on the large Cu $3d_{x^2-y^2}$ sheets and weaker on the small pockets around $M$. These results show that, once the Cu magnetic moments vanish, the EPC arises from Cu--O bond-stretching and mixed Cu--O modes between 30 and 70~meV, as in the undoped nonmagnetic phase (SM); neither the rotational oxygen modes nor the low-frequency phonons contribute appreciably.
\par}

{\subsection{Doping dependence of EPC and the superconducting dome }
\label{sec:doping_tc}

Table~\ref{tab:doping_tc} summarizes the doping dependence of $\lambda$ and the estimated $T_{\mathrm c}$. The coupling increases from $\lambda=0.37$ at 12.5\% to 0.47 at 25\% and 1.57 at 37.5\%, the largest value among the sampled dopings, and decreases to 0.42 at 56.25\%, where the Cu magnetic moments vanish. The phonons that dominate the coupling also change with doping. At 25\%, the rotational oxygen modes dominate the coupling in the range of 50--65~meV (Sec.~\ref{sec:epc_25}). At 37.5\%, these rotational modes still couple strongly, and phonons below 45~meV add a large contribution through stronger coupling near the Fermi-surface nesting wave vector, the larger $N(E_F)$, and the phonon softening (Sec.~\ref{sec:epc_375}). At 56.25\%, the coupling comes from Cu--O bond-stretching and mixed Cu--O modes between 30 and 70~meV, and neither the rotational oxygen modes nor the low-frequency phonons contribute appreciably (Sec.~\ref{sec:epc_5625}). Because only four dopings are sampled, the doping of maximal $\lambda$ between 25\% and 56.25\% is not determined. However, the EPC-superconducting dome terminates close to the point where a VHS crosses the Fermi level (below 56.25\% doping, Fig. 1(f)) -- a property also found in experimental cuprate superconducting domes~\cite{Scalapino}.

\begin{table}[ht]
\caption{\textbf{Doping dependence of the EPC strength and the estimated superconducting transition temperature.} EPC strength $\lambda$, logarithmic phonon energy $E_{\log}$, and $T_{\mathrm c}$ estimated from Eq.~(\ref{eq:tc_doping}) with $\mu^*=0.10$ and $\mu^*=0.04$ for hole-doped CaCuO$_2$. The states at 12.5\%, 25\%, and 37.5\% are $G$-AFM; the state at 56.25\% is nonmagnetic. $T_{\mathrm c}$ includes only phonon-mediated pairing.}
\label{tab:doping_tc}
\centering
\renewcommand{\arraystretch}{1.3}
\begin{tabular*}{\columnwidth}{@{\extracolsep{\fill}}ccccc}
\hline\hline
Hole doping & $\lambda$ & $\omega_{\log}$ & \multicolumn{2}{c}{$T_{\mathrm c}$ (K)} \\
(\%) & & (meV) & $\mu^*=0.10$ & $\mu^*=0.04$ \\
\hline
12.5 & 0.368 & 52.85 & 1.54 &5.88 \\
25 & 0.472 & 57.01 & 6.31 & 14.40 \\
37.5 & 1.569 & 31.22 & 43.05 & 50.25 \\
56.25 & 0.423 & 38.23 & 2.53 & 6.97 \\
\hline\hline
\end{tabular*}
\end{table}

The logarithmic phonon energy $\omega_{\log}$ is the geometric mean of phonon energies weighted by their contributions to $\lambda$ (SM Sec.~S-I). 
It increases slightly from 52.9~meV at 12.5\% to 57.0~meV at 25\%, decreases to 31.2~meV at 37.5\% as the low-frequency coupling grows, and recovers to 38.2~meV at 56.25\%, where this coupling is absent.

We estimate $T_{\mathrm c}$ using the isotropic McMillan--Allen--Dynes expression~\cite{FelicianoRMP2017},
\begin{equation}
T_{\mathrm c}=\frac{\omega_{\log}}{1.2k_{\mathrm B}}
\exp\!\left[-\frac{1.04(1+\lambda)}{\lambda-\mu^*(1+0.62\lambda)}\right],
\label{eq:tc_doping}
\end{equation}
with the Coulomb pseudopotential $\mu^*$ fixed across dopings. From 25\% to 37.5\%, the increase of $\lambda$ outweighs the decrease of $E_{\log}$ and raises $T_{\mathrm c}$ from 6--14~K to 43--50~K for $\mu^*$ between 0.10 and 0.04. At 56.25\%, the decrease of $\lambda$ outweighs the recovery of $\omega_{\log}$ and lowers $T_{\mathrm c}$ to 2.5--7.0~K.  
\par}

{\subsection{Rotational phonons and the thermal Hall effect}
\label{sec:thermal_hall}

\begin{figure*}[ht]\centering
\includegraphics[width=\textwidth]{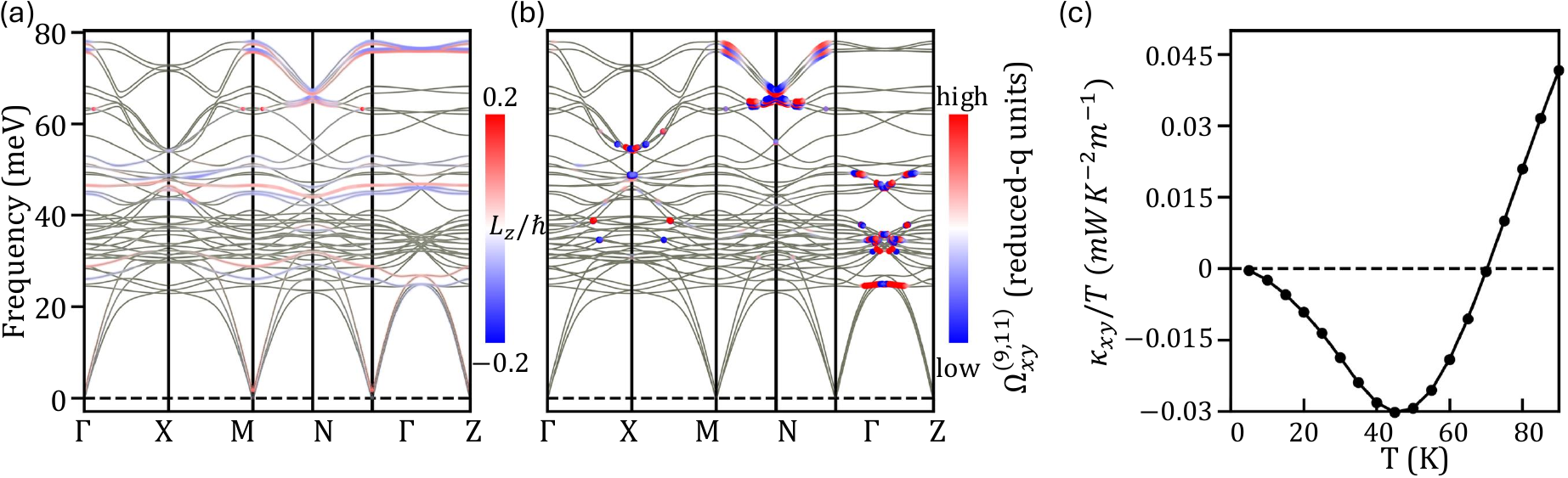}
\caption{\textbf{Phonon angular momentum, Berry curvature, and thermal Hall conductivity with the gyrotropic term equivalent to an external magnetic field of 15 T, at 25\% doping.} (a) Phonon dispersion colored by the $z$ component of the phonon angular momentum $L_z$ (in units of $\hbar$) of the oxygen atoms O$_1$ and O$_3$ [numbered as in Fig.~S3(c) of the SM]. (b) Phonon dispersion colored by the Berry curvature $\Omega_{xy}$ (in reduced-$\mathbf q$ units); the marker size scales with $|\Omega_{xy}|$. (c) Intrinsic phonon thermal Hall conductivity $\kappa_{xy}/T$ as a function of temperature.}
\label{fig:thermal_hall}
\end{figure*}

The phonon thermal Hall response observed in cuprates~\cite{TaillThPh} motivates examining whether the rotational modes identified here can support transverse heat transport.
In the unperturbed lattice, symmetry-related oxygen sites circulate in opposite directions and thus carry opposing local phonon angular momenta~\cite{Das2026AFCP}.
Their circular motion can also produce orbital magnetic moments through the ionic Born effective charges~\cite{Juraschek2019PhononMoments}, but these moments are staggered and cancel when integrated over the unit cell.
The unperturbed harmonic lattice dynamics retain an effective time-reversal symmetry, enforcing $\Omega_{\nu,xy}(\mathbf q)=-\Omega_{\nu,xy}(-\mathbf q)$ and causing contributions from opposite wave vectors to cancel. With inversion symmetry as well, the phonon Berry curvature $\Omega_{\nu,xy}(\mathbf q)$ vanishes for an isolated nondegenerate phonon mode.
Local circular polarization therefore does not by itself establish a finite intrinsic thermal Hall conductivity.
Compensated angular momentum, vanishing at every site, has also been found in symmetric lattices without magnetic order~\cite{Yi2026HiddenAM}. Here the strongly coupled rotational modes appear only in the doped magnetic state.

To obtain a finite thermal Hall conductivity in CCO, TRS between the symmetry-related oxygen sites must be lifted. To examine how this cancellation can be lifted, we introduce a phenomenological gyrotropic interaction on the Cu degrees of freedom,
\begin{equation}
 \left[D(\mathbf q)-\omega_{n\mathbf q}^{2}I
 -2i\omega_{n\mathbf q}G{(r)}\right]
 \mathbf e_{n\mathbf q}=0,
 \label{eq:gyro}
\end{equation}
Here, $D(\mathbf q)$ is the unperturbed mass-weighted dynamical matrix, $G(r)$ is a real antisymmetric operator acting on the phonon displacements, and $\mathbf e_{\nu\mathbf q}$ is the phonon eigenvector.
The velocity-dependent term also inherits a parameter $r$ that represents the action of an external magnetic field or a magnetic environment on the lattice. 
It breaks the effective time-reversal symmetry of the phonon problem and distinguishes modes of opposite handedness. 
However, in the nonmagnetic configuration of CCO, where the circular motion of the oxygen atoms is absent (see Fig. S5), a finite net Berry curvature from the gyrotropic term is not expected.

Figure~\ref{fig:thermal_hall} shows the results for the illustrative parameter $r$=0.42 THz, equivalent to an external magnetic field of 15T (SM Section S.XI). The gyrotropic term modifies the coupled copper-oxygen motion. For example, the two oxygen atoms that are not related by inversion symmetry exhibit finite phonon angular momentum $L_z$, primarily on branches between 42 and 66~meV and on the highest-energy branch along $\Gamma$--$Z$ [Fig.~\ref{fig:thermal_hall}(a)]. The perturbation thus produces collective Cu--O rotational modes with nonzero Berry curvature. The Berry curvature $\Omega_{xy}$ is large near $M$ between 48 and 57~meV, on the branch near 70~meV along $X$--$M$--$N$, near $\Gamma$ at $\sim$77~meV, and on the degenerate acoustic phonon branches between 22 and 26~meV along $\Gamma$--$Z$ [Fig.~\ref{fig:thermal_hall}(b)]. It therefore arises both in the low-frequency region and in the region containing the rotational oxygen modes. 
The thermal Hall response consequently depends on the collective Cu-O eigenmodes; the dominant oxygen contribution to EPC does not imply a dominant oxygen contribution to thermal Hall transport.

Within the intrinsic phonon Hall formalism of Ref.~\cite{Qin2012PhononHall}, $\kappa_{xy}$ is determined by the BZ integral of the mode-resolved Berry curvature with thermal occupation weights and the energy-magnetization correction. 
Removing the symmetry-enforced cancellation allows a finite intrinsic response [Fig.~\ref{fig:thermal_hall}(c)]. The calculated $\kappa_{xy}/T$ is negative below $\sim$66~K, with a minimum of about $-0.027$~mW\,K$^{-2}$\,m$^{-1}$ near 45~K, changes sign near 66~K, and reaches 0.049~mW\,K$^{-2}$\,m$^{-1}$ at 90~K. The sign change shows that branches at different energies contribute Berry curvature of opposite sign, and their relative thermal weight changes with temperature. 
The gyrotropic model thus provides a way for rotational modes to contribute to thermal Hall transport, although comparing it quantitatively with experiment requires accounting for contributions from different carriers. An alternative mechanism for THE is provided by phonons scattering asymmetrically off fluctuating spin structures in the pseudogap~\cite{Guo2022}; our strongly coupled rotational modes are precisely the phonons most susceptible to such scattering. 

As shown in  Fig.~\ref{fig:thermal_hall}(c), up to 60~K, the shape of $\kappa_{xy}/T$ is qualitatively similar to the form expected for acoustic phonons~\cite{Yang2020UniversalTHE}. However, the experimental $\kappa_{xy}$ for cuprates~\cite{Grissonnanche2019GiantTHE,TaillThPh} is negative at all temperatures and falls smoothly to zero as $T$ increases, whereas the calculated value changes sign near 66~K. Moreover, the magnitude and the temperature scale differ: the experimental peak of $\kappa_{xy}/T$ is about 50 times larger, and it occurs at a temperature about four times lower. This is consistent with many THE experiments, which find that the experimental $\kappa_{xy}$ is much larger than expected for the applied magnetic field.

\par}

\section{Discussion}
The rotational modes identified here are relevant to the thermal Hall effect observed in the pseudogap regime of cuprates, which vanishes when the pseudogap closes and has been proposed to originate from phonons carrying angular momentum~\cite{TaillThPh}. In our calculations, the rotational modes with strong EPC emerge only in the doped magnetic phase and are absent in the nonmagnetic phase [Fig.~S5(e) of SM], and the gyrotropic model (Sec.~\ref{sec:thermal_hall}) shows that they can produce a finite phonon thermal Hall conductivity~\cite{Taill1,Tranq_chir,Ramshaw2026}. If these modes are the source of the measured signal, this would support the interpretation of the pseudogap as a regime of short-range antiferromagnetic order~\cite{Yaohong_PRL_2021}. A direct test is offered by resonant inelastic x-ray scattering (RIXS) with circularly polarized photons, which can detect phonon angular momentum~\cite{Ueda2023}.

Strong EPC, when combined with strong electronic correlations and/or antiferromagnetic fluctuations, has long been proposed as a mechanism of high-temperature superconductivity with $d$-wave pairing symmetry~\cite{KULIC20001,yaohong_sb}. Although it remains debated whether isotropic EPC alone can lead to $d$-wave pairing~\cite{dwave1,dwave2}, Ref.~\cite{KULIC20001} shows that the effect is strongest for highly anisotropic EPC dominated by forward scattering. In this context, recent DFT+$U$ calculations have reported strong EPC in the parent compound La$_2$CuO$_4$ when local magnetism is included~\cite{DFTU_EPC_LCO}, although not associated with rotational phonons (see also SM Section S-III). Furthermore, a model study suggests that explaining the low-energy kinks observed in cuprates likely requires the intertwined effects of both EPC and electronic correlations~\cite{Huang_epc_correlation}.

Notably, it was recently shown that the bond Su--Schrieffer--Heeger (SSH) phonons---that is, phonons that modulate the electronic hopping amplitude $t$---can cooperate with strong correlations to produce $d$-wave superconductivity in a phase characterized by strong antiferromagnetic fluctuations~\cite{yaohong_sb,Yaohong_PRL_2021}.
Our study thus contributes a new element to this emerging picture: in cuprates, as well as in the nickelates studies in Ref.~\cite{ZhangPRLLaNiO2,Zhang_LNO_EPC}, strong (and fluctuating) antiferromagnetic order itself generates enhanced EPC. Interestingly, this occurs even though different phonons dominate the coupling in the two materials. In nickelates, the dominant EPC arises from low-energy La/Ni modes that couple strongly to a flat electronic band of $d_{z^2}$ character~\cite{ZhangPRLLaNiO2,Zhang_LNO_EPC}, while rotational oxygen phonons play an important role in cuprates. This points to magnetism-enhanced EPC as a unifying mechanism across different families of unconventional superconductors, despite substantial differences in their lattice dynamics and electronic structure. In our calculations, the rotational oxygen modes modulate the Cu--O hopping and couple strongly to the Cu $3d_{x^2-y^2}$ electrons only in the presence of antiferromagnetic correlations. They may therefore act together with spin fluctuations in cuprate pairing~\cite{Yaohong_PRL_2021,yaohong_sb}. Angular-momentum-carrying phonons have been proposed in a model to enhance spin-fluctuation pairing~\cite{Gao2023ChiralPhononSC}. However, our collinear calculations without spin--orbit coupling cannot test this mechanism. Our calculations also show a Fermi-surface nesting peak and a phonon softening near the experimental CDW wave vector (Sec.~\ref{sec:epc_375} and SM Sec.~S-X), without establishing a CDW instability. RIXS on Hg1223 reports a bond-stretching-phonon anomaly and a paramagnon softening at $\mathbf q\approx(0.30,0)$~\cite{Rongrueangkul2026Hg1223}; this is a consistency check, not a test of the rotational modes.

\section{Conclusion}
Our study establishes a mode-resolved framework, based on the SCAN functional without a Hubbard $U$, for quantifying EPC in cuprates, with the key finding that magnetism dramatically enhances the EPC. At 25\% doping, the resulting strong EPC is governed primarily by rotational (circularly polarized) oxygen phonons rather than the conventionally emphasized oxygen breathing modes, providing a new microscopic basis for interpreting ubiquitous self-energy anomalies. With increasing doping, $\lambda$ increases to $\sim$1.6 at 37.5\%, where low-frequency phonons soften and couple strongly near the Fermi-surface nesting wave vector, and decreases to $\sim$0.4 at 56.25\%, where the Cu magnetic moments vanish; the strong EPC thus follows the presence of the Cu magnetic moments. The rotational modes carry staggered angular momentum, and a time-reversal-breaking coupling on the Cu sites lifts its cancellation and yields a finite phonon thermal Hall conductivity. In sum, our results identify \emph{magnetism-induced EPC}, mediated by rotational oxygen phonons and, at higher doping, by low-frequency phonons, as a key ingredient linking the pseudogap, anomalous thermal transport, and superconductivity in cuprates, and suggests a unifying mechanism that may extend to other strongly correlated materials.

\bibliography{Ref_wo_url,Ref_revA_new}

\SB{Acknowledgements}\textbf{\---}We thank Dr. Hongtao Yan and Prof. X.J. Zhou for helpful discussions on the kinks of cuprates. R.Z. also thanks Prof. Lu Chen for insightful discussions on the thermal Hall effects of cuprates. Y.W., R.Z., J.P.P. and J.S. acknowledge the support of the U.S. National Science Foundation under Grant No. CHE-2533416. A.R. acknowledges the U.S. National Science Foundation under Grant No. CHE-2533416 for the meta-GGA work. For the work on superconductors with electron-phonon interaction, A.R. acknowledges the U.S. Department of Energy, Office of Science, Office of Basic Energy Sciences, under Award Number DE-SC0026293. R.Z. and J.S. also acknowledge the support of the American Chemical Society Petroleum Research Fund, grant number: PRF\#69244-ND10. R.Z. and J.S. were supported in part by the U.S. Department of Energy’s Genesis Mission through under Agreement No. DE-SC0026652. The work at Tulane University was also supported by the Advanced Cyberinfrastructure Coordination Ecosystem funded by the National Science Foundation, and the National Energy Research Scientific Computing Center (NERSC) using NERSC Awards No. BES-ERCAP0020494 and No. BESERCAP0028709. The work at Northeastern University was supported by the National Science Foundation through the Expand-QISE award NSF-OMA-2329067 and benefited from the resources of Northeastern University’s Advanced Scientific Computation Center, the Discovery Cluster, and the Massachusetts Technology Collaborative award MTC-22032. The work at Los Alamos National Laboratory was carried out under the auspices of the U.S. Department of Energy (DOE) National Nuclear Security Administration under Contract No. 89233218CNA000001. It was supported by the LANL LDRD Program, the Quantum Science Center, a U.S. DOE Office of Science National Quantum Information Science Research Center, and in part by the Center for Integrated Nanotechnologies, a DOE BES user facility, in partnership with the LANL Institutional Computing Program for computational resources. Additional computations were performed at the NERSC, a U.S. Department of Energy Office of Science User
Facility located at Lawrence Berkeley National Laboratory, operated under Contract No. DE-AC02-05CH11231 using NERSC Award No. ERCAP0020494. B.B. acknowledges support from COST Action CA16218. B.S. acknowledges the support of the Department of Atomic Energy, Government of India, under Project Identification Nos. RTI4013 and RTI4015. Howard University was supported by the National Science Foundation through the Expand-QISE award
NSF-OMA-2329067. Research at Howard University utilized computational resources provided through ACCESS allocation PHY260045.

\SB{Author contributions}\textbf{\---}R.Z., J.S., and A.B. designed the study. Y.W., M.E., and R.Z. performed first-principles computations and analyzed the data, with help from C.L., H.M., L.H, B.B., R.S.M., A.R., J.P., A.B., and J.S.; S.M., B.S., and S.C. performed the calculations of the thermal Hall effect. A.B. and J.S. provided computational infrastructure. R.Z., Y.W, C.L., S.M., B.S., B.B, R.S.M., A.B., and J.S. wrote the manuscript with input from all the authors. All authors contributed to editing the manuscript.

\SB{Additional information}\textbf{\---}The authors declare no competing financial interests.

\clearpage
\onecolumngrid
\begin{center}
{\large\textbf{Supplemental Material for ``Rotational Phonons Activated by Magnetism as a Key Ingredient of Cuprate Physics''}}
\end{center}
\vspace{1em}
\twocolumngrid
\beginsupplement
\section{Computational details}

All the calculations were performed using the projector-augmented-wave (PAW) method~\cite{Kresse1999}, as implemented in the Vienna ab initio simulation package (VASP)~\cite{Kresse1993,Kresse1996}. Since our previous studies have shown that the SCAN density functional gives an accurate description of the electronic and phonon structures of cuprates~\cite{Furness2018,Lane2018,ZhangYPANS,NingPRBYBCO6,NingJCPDFAYBCO6}, we use SCAN for the EPC study of CaCuO$_2$ (CCO), to remain consistent with our earlier work. A high-energy cutoff of 520 eV was applied to truncate the plane-wave basis set. Crystal structures and ionic positions were fully optimized with a force convergence criterion of 0.001 eV/\AA{} at each atom and a total energy tolerance of $10^{-8}$ eV. For the nonmagnetic (NM) configuration of CCO, a $\Gamma$-centered $k$-point mesh of $15 \times 15 \times 12$ was employed for structural relaxation. Phonon calculations were carried out using a $4 \times 4 \times 4$ supercell with a $\Gamma$-centered k-point mesh of $3 \times 3 \times 3$, employing the finite displacement method. A Gaussian smearing with a small $\sigma$ of 0.2 eV was applied in the phonon calculations. Computations for the magnetic $G$-AFM phase employed: a $\sqrt{2} \times \sqrt{2} \times 2$ supercell; a $\Gamma$-centered $11 \times 11 \times 9$ $k$ point mesh for structural relaxation; and, a $3 \times 3 \times 3$ supercell of the $G$-AFM unit cell with a $4 \times 4 \times 3$ $\Gamma$-centered $k$-point mesh for phonon calculations based on the finite displacement method. Cu 3$d$ and O 2$p$ orbitals were included in constructing the Wannier model. A non-integer nuclear charge approach was used to model effects of hole doping by assigning a fractional nuclear charge to the Ca site~\cite{NNC_1D}. The corresponding Ca pseudopotentials with non-integral $Z$ values were generated with the VASP code. All calculations were collinear and without spin--orbit coupling.

For EPC calculations, we followed the Wannier-interpolation procedure outlined in Ref.~\cite{ManuelPRB2020} with the all-electron EPC matrix elements defined in Ref.~\cite{ManuelPRB2022,FDEPCPAW}. The Wannier representation of the electronic band structure was obtained using Wannier90~\cite{MOSTOFI20142309} with the help of the VASP2WANNIER90 interface. For the undoped insulating $G$-AFM phase, the long-range (Fr\"ohlich) interaction was included through the nonanalytic correction, which produces the LO--TO splitting at $\Gamma$ [Fig.~\ref{fig:cco_afm}(b)]. For the doped metallic phases, the free carriers screen the long-range interaction, and no nonanalytic correction was applied. The EPC strength $\lambda$ and the isotropic Eliashberg spectral function $\alpha^{2}F$ were computed based on the following equations.

Within the Migdal approximation~\cite{Migdal1958}, the imaginary part of the phonon self-energy is expressed as
\begin{equation}\label{eq:phse}
\Pi''_{\vecq\nu} = \text{Im} \sum_{m,n,\veck} w_{\veck}  \vert g_{mn,\nu}( {\veck,\vecq} ) \vert^2 
\frac{ f_{n\veck} - f_{m{\veck+\vecq}} }{
\epsilon_{m{\veck+\vecq}} - \epsilon_{n\veck} - \omega_{\vecq\nu} - i\eta},
\end{equation}
where $f_{n{\veck}}$ is the Fermi occupation of the single-particle state $\vert\psi_{n\veck}\rangle$ associated with the eigenvalue $\epsilon_{n\veck}$, $w_{\veck}$ are the weights of the dense ${\veck}$-points, and $\eta$ is a real positive infinitesimal. Spin index is omitted for simplicity. The total EPC strength $\lambda$ is given by the full BZ average of the mode-resolved contributions $\lambda_{\vecq\nu}$:
\begin{equation}\label{eq:lam}
\lambda = \sum_{\vecq\nu} w_{\vecq} \lambda_{\vecq\nu},
\end{equation}
where $w_{\vecq}$ are the weights of the dense $\vecq$-points. The double delta approximation~\cite{epiq_impl2024} was adopted for calculating the imaginary part of the phonon self-energy, under which Eq.~\eqref{eq:phse} can be rewritten as follows:

\begin{equation}\label{eq:phse_doubledelta}
\Pi''_{\vecq\nu} = \sum_{m,n,\veck} w_{\veck} \omega_{\vecq\nu} \vert g_{mn,\nu}( {\veck,\vecq} ) \vert^2 \delta(\epsilon_{n{\bf k}})\delta(\epsilon_{m\veck+\vecq})
\end{equation}
$\lambda_{\vecq\nu}$ can now be rewritten as:
\begin{equation}
\lambda_{\vecq\nu} = \frac{1}{\pi N_\text{F}} \frac{ \Pi''_{\vecq\nu}}{ \omega^2_{\vecq\nu}},
\end{equation}
where $N_\text{F}$ is the density of states at the Fermi level. Using this mode-resolved $\lambda_{\vecq\nu}$ and the phonon frequencies $\omega_{\vecq\nu}$, the isotropic Eliashberg spectral function, $\alpha^2 F$, can be expressed as:
\begin{equation}\label{eq:a2F}
\alpha^2F(\omega) = \frac{1}{2}\sum_{\vecq\nu}  
w_{\vecq} \omega_{\vecq\nu} \lambda_{\vecq\nu} \, \delta( \omega - \omega_{\vecq\nu}).
\end{equation}

The phonon-mediated superconducting transition temperature, $T_\text{c}$, was evaluated using the McMillian-Allen-Dynes formula~\cite{McMillan1968},
\begin{equation}
T_\text{c} = \frac{\omega_\text{log}}{1.2}\exp\left[{\frac{-1.04(1+\lambda)}{\lambda(1 - 0.62\mu^*) - \mu^*}}\right],
\label{eq:tc}
\end{equation}
where the logarithmic average frequency $\omega_{\text{log}}$ can be calculated from the following equation~\cite{Allen1975,grimvall1981electron}
\begin{equation}\label{wlog}
\omega_\text{log} = \exp\left[\frac{2}{\lambda}\int^\infty_0\text{d}\omega \frac{ \log\,\omega}{\omega}\alpha^2F(\omega)\right].
\end{equation}

The Fan-Migdal term, which accounts for the EPC contributions to the electronic self-energy was evaluated by using:
\begin{align}
\Sigma^\text{FM}_{n\veck}(\omega,\, T) ={}& \sum_m\sum_{\vecq\nu} w_{\vecq} \left\vert g_{mn,\nu}(\veck,\,\vecq)\right\vert^2 \nonumber\\
&\times\left[ \frac{ 1- f_{m\veck+\vecq}(T) + n_{\vecq\nu}(T) }{ \omega -\epsilon_{m\veck+\vecq} -\omega_{\vecq\nu} + i\eta} \right. \nonumber\\
&\left. + \frac{ f_{m\veck+\vecq}(T) + n_{\vecq\nu}(T) }{ \omega -\epsilon_{m\veck+\vecq} +\omega_{\vecq\nu} + i\eta} \right],
\label{FM-self}
\end{align}
where $f_{m\veck+\vecq}(T)$ and $n_{\vecq\nu}(T)$ are the Fermi-Dirac distribution function and Bose-Einstein distribution function for electrons and phonons of the system at the finite temperature $T$, respectively. The infinitesimal positive quantity $\eta$ here ensures causality and serves as an empirical tuning parameter in practical calculations~\cite{Gonze_Marini2015/JCP}. 

An accurate evaluation of the electron self-energy from EPC requires the inclusion of both the Fan-Migdal (FM) and Debye-Waller (DW) terms~\cite{FelicianoRMP2017}. However, converging the real part of the self-energy requires summing over a large number of empty states, which is not practical within the Wannier basis. Therefore, for pristine AFM cuprates, we set the real part of the self-energy to zero and focused only on estimating the spectral function broadening in our calculations via the equation: 

\begin{align}
A_{\veck}(\omega) &= -\frac{1}{\pi} \sum_{n} \cfrac{ \text{Im}\Sigma^\text{e-ph}_{n\veck}(\omega) }{ \big(\omega - \epsilon_{n\veck}  \big)^2 +\left[ \text{Im}\Sigma^\text{e-ph}_{n\veck}(\omega) \right]^2}
\label{Akw_pristine}
\end{align}

For pristine cuprates with non-magnetic configuration and doped case, we can subtract $\text{Re}\Sigma^\text{e-ph}_{n\veck}(E_F)$ from the calculated $\text{Re}\Sigma^\text{e-ph}_{n\veck}(\omega)$ to ensure electron number conservation~\cite{FelicianoRMP2017,epwcpc} according to Luttinger's theorem~\cite{Luttinger_1960} as follows:

\begin{equation}
\mathrm{Re}\,\overline{\Sigma}^{\mathrm{e\text{-}ph}}_{n\mathbf{k}}(\omega)
=
\mathrm{Re}\,\Sigma^{\mathrm{e\text{-}ph}}_{n\mathbf{k}}(\omega)
-
\mathrm{Re}\,\Sigma^{\mathrm{e\text{-}ph}}_{n\mathbf{k}}(E_F)\,
\end{equation}

The electronic spectral function is then obtained from the calculated phonon-induced self-energy using:
\begin{align}
	A_{\veck}(\omega) ={}& -\frac{1}{\pi} \sum_{n} \text{Im}\Sigma^\text{e-ph}_{n\veck}(\omega) \Big\{ \big[ \omega - \epsilon_{n\veck} -\text{Re}\overline{\Sigma}^\text{e-ph}_{n\veck}(\omega) \big]^2 \nonumber\\
	& +\big[ \text{Im}\Sigma^\text{e-ph}_{n\veck}(\omega) \big]^2 \Big\}^{-1}
	\label{Akw}
\end{align}

The fermi-nesting function, which describes electron-hole excitations around the fermi surface, is evaluated by using
\begin{align}
	\chi_\text{nest}(\boldsymbol{q}) &= \sum_{m,n,\boldsymbol{k}} \delta(\epsilon_{n\boldsymbol{k}} - \epsilon_F) \delta( \epsilon_{m{\boldsymbol{k+q}}} -\epsilon_F ).
\end{align}

\section{EPC in pristine C\MakeLowercase{a}C\MakeLowercase{u}O$_2$ with NM configuration}

\begin{figure*}[t!]
\centering
\includegraphics[width=.9\textwidth]{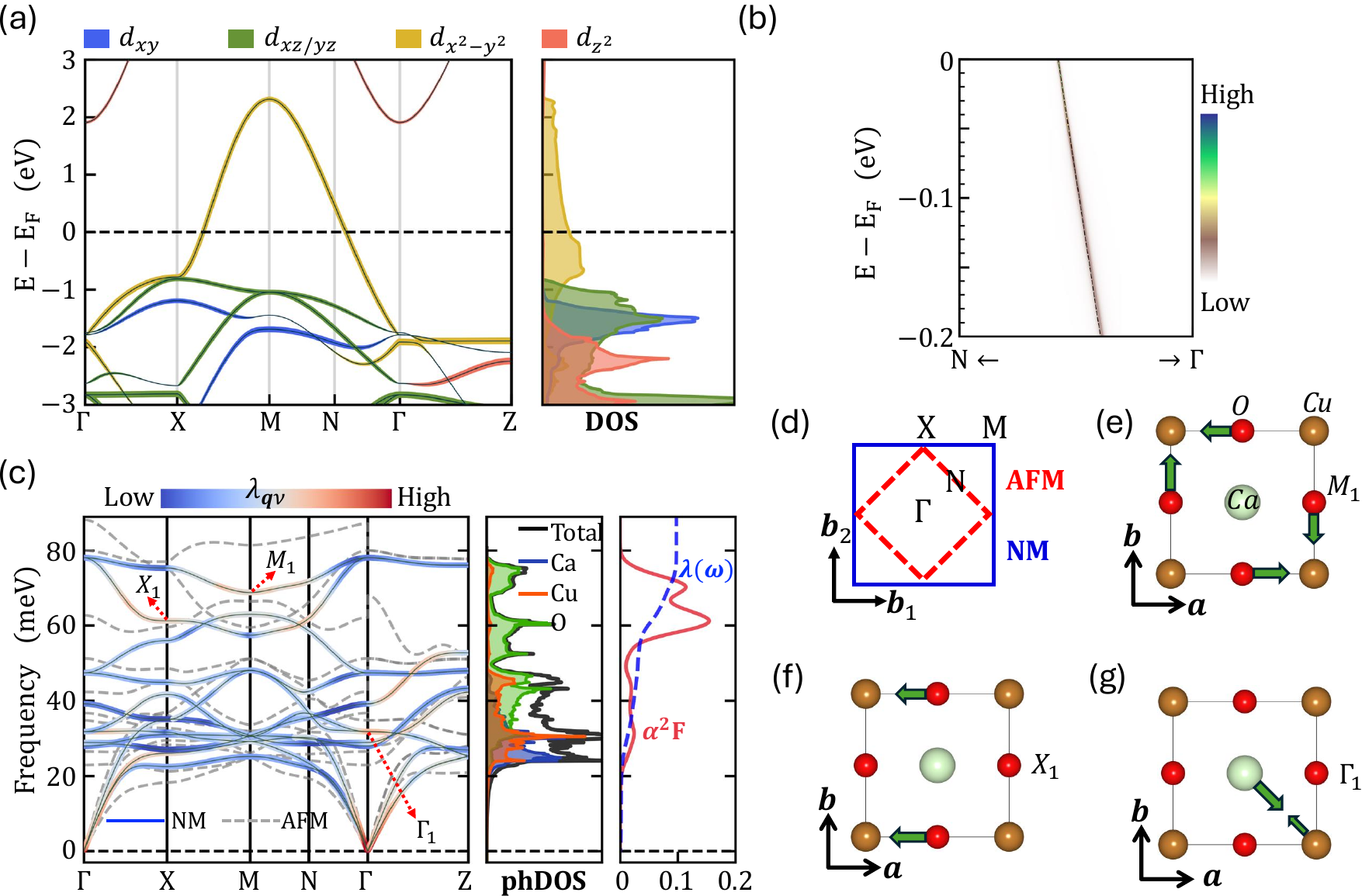}
\caption{\textbf{Calculated electronic structure, phonon dispersion, and EPC of NM phase of pristine CCO}. (a) Band structure along the high-symmetry lines in the first BZ. Contributions of various Cu $d$-orbitals are shown. High-symmetry points $\Gamma$, X, and M lie in the $k_z =0$ plane, while point Z lies in the $k_z = \pi/c$ plane. (b) Spectral function along the $\Gamma$–$N$ line. (c) Left panel: phonon dispersions (solid lines) along the high-symmetry lines in the first BZ; marker color-intensity indicates EPC strength, $\lambda_{\textbf{q}\nu}$. Phonon dispersions of the $G$-AFM phase (dashed gray lines, unfolded in the first NM BZ) are also shown; Middle panel: site-projected phonon density of states (PhDOS); Right panel: Eliashberg spectral functions, $\alpha^2F(\omega)$ (solid red lines), and the total EPC strength, $\lambda$ (dashed blue lines). A dense $k$-mesh of $60 \times 60 \times 60$ and a $q$-mesh of $30 \times 30 \times 30$ were used for accurate BZ sampling. (d) Schematic of the NM and AFM BZs in the $k_z = 0$ plane, with high-symmetry points marked. (e)-(g) Side views of selected phonon modes marked in (c), with green arrows representing the direction of atomic vibrations. Brown, green, and red balls represent Cu, Ca, and O atoms, respectively.}
\label{fig:cco_nm}
\end{figure*}

We begin by investigating EPC in the NM configuration of pristine CaCuO$_2$. Figure~\ref{fig:cco_nm} (a) shows the electronic structure, where the states at the Fermi level are dominated by the Cu 3$d_{x^2-y^2}$ orbitals. Notably, a van Hove singularity appears near the X point at –0.7 eV, which is further confirmed by the DOS plot shown in the right panel of Fig.~\ref{fig:cco_nm} (a). The corresponding phonon dispersions (solid lines) are displayed in the left panel of Fig.~\ref{fig:cco_nm} (c), with the phonon density of states (PhDOS) presented in the middle panel. High-frequency optical modes above 50 meV originate primarily from oxygen vibrations, whereas low-frequency modes are mainly associated with motions of Cu and Ca atoms.

Mode-resolved EPC in the left panel of Fig.~\ref{fig:cco_nm}(c) shows that the full- and half-breathing oxygen modes make leading contributions to the total EPC, producing prominent peaks in the $\alpha^{2}F(\omega)$ [right panel of Fig.~\ref{fig:cco_nm}(c)]. Low-frequency vibrations of Cu and Ca atoms yield a relatively smaller peak in $\alpha^{2}F(\omega)$. Multiple phonon branches thus contribute to the EPC in the NM phase of CCO, even though the total EPC strength is dominated by the O-breathing modes; representative eigenmodes that play a key role are depicted in Fig.~\ref{fig:cco_nm}(e–g). However, the total predicted EPC strength in the NM phase is small, with $\lambda = 0.1$, yielding $T_c$ = 0 for $\mu*$ = 0.1. In line with these results, the NM spectral function in Fig.~\ref{fig:cco_nm}(b) shows no kinks. The weak EPC found in the NM phase from the present SCAN calculations is in line with previous DFT studies~\cite{giustino2008cuprate}. 

We emphasize that the NM configuration is not relevant for describing the physics of cuprates. For the pristine case, NM-DFT yields a metallic CCO, in sharp disagreement with the experimentally observed AFM insulator state~\cite{Furness2018}. In fact, the NM phase is energetically unstable in our SCAN calculations, lying 73.5 meV/Cu above the $ G$-type AFM ($G$-AFM) configuration. Moreover, the unfolded phonon dispersions of the $G$-AFM phase [gray lines in Fig.~\ref{fig:cco_nm}(c)] differ substantially from the NM solution, indicating that magnetic effects play a critical role in shaping the EPC in cuprates. 

\section{EPC in pristine C\MakeLowercase{a}C\MakeLowercase{u}O$_2$ with $G$-AFM configuration}

\begin{figure*}[t!]
\centering
\includegraphics[width=.85\textwidth]{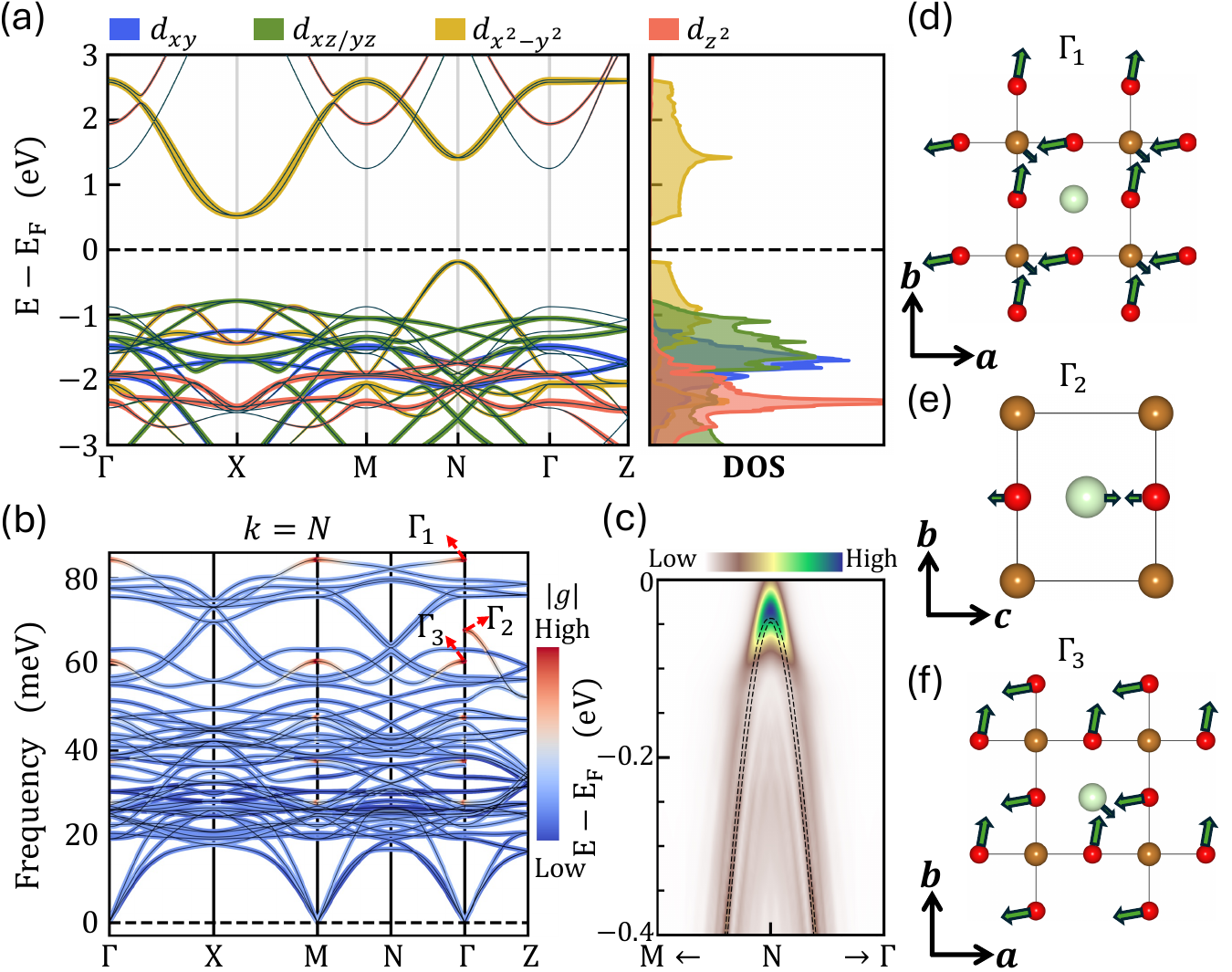}
\caption{\textbf{Calculated electronic structure, phonon dispersion, and EPC for $G$-type AFM phase of CCO}. (a) Orbital-projected band structure along the high-symmetry lines in the BZ. Contributions of various Cu $d$-orbitals are shown; side panel shows the corresponding partial densities of states. (b) Folded phonon dispersion along the high-symmetry lines in the BZ. Marker colors along the $\mathbf{q}$-path show the magnitude of the EPC matrix elements, calculated by summing over bands ($N_b$) selected to include Cu 3$d$ orbitals: $|g^\sigma_{\vecq\nu}| \equiv \left( \sum_{mn} |g_{mn,\nu}^\sigma(\veck=K,\vecq)|^2/N_{b} \right)^{1/2}$, where $\sigma$ is chosen to be the spin-up channel, $\vecq$ follows the high-symmetry lines in the bulk BZ, and $\veck$ is fixed at the $N$ point. (c) Spectral function at the VBM. Dashed blue line gives the bare DFT bands. (d)-(f) Side view of selected phonon modes marked in (b), with green arrows giving directions of atomic vibrations. Brown, green, and red balls represent Cu, Ca, and O atoms, respectively.}
\label{fig:cco_afm}
\end{figure*}

Figure~\ref{fig:cco_afm}(a) shows the calculated electronic structure of the $G$-AFM phase. We find that the Cu 3$d_{x^2-y^2}$ bands are fully gapped, with an indirect band gap of approximately 0.5 eV. Both the conduction-band minimum (CBM) and valence-band maximum (VBM) are primarily composed of Cu 3$d_{x^2-y^2}$ states. Interestingly, VHSs are observed near both the VBM and CBM [see DOS plot in the right-hand panel of Fig.~\ref{fig:cco_afm}(a)], which can enhance many-body interactions in the cuprates. The unfolded phonon dispersions of the $G$-AFM phase (dashed gray lines) are shown in Fig.~\ref{fig:cco_nm}(c). Compared to the NM case, the optical modes in the $G$-AFM phase exhibit noticeable hardening—particularly the oxygen breathing modes—while the low-frequency modes soften. These findings suggest that Cu–O interactions are enhanced in the magnetic phase relative to the NM phase. Additionally, by including non-analytical corrections, clear LO–TO mode splittings at the $\Gamma$ point are found in the phonon dispersions of the $G$-AFM phase, as is the case in CoO, NiO, and BaTiO$_3$~\cite{wang_epc_rrscan}.

To explore the EPC in the $G$-AFM phase of CCO, we present the folded phonon dispersions with projected $g$-matrix elements in Fig.~\ref{fig:cco_afm}(b). Strong interactions between electrons and the LO–TO phonon modes, driven by the Fröhlich mechanism—are observed in this magnetic phase, with the maximum $g$-matrix elements exceeding $10^4$ meV near the $\Gamma$ point [see Fig.~\ref{fig:cco_afm}(b)]. Beyond the well-known oxygen breathing modes, we also identify several LO and TO modes [marked in Fig.~\ref{fig:cco_afm}(b)] that contribute significantly to the total EPC. For example, the LO $\Gamma_{1}$ [Fig.~\ref{fig:cco_afm}(d)] and $\Gamma_{3}$ [Fig.~\ref{fig:cco_afm}(f)] modes involve in-plane Cu–O bond bending, whereas the TO $\Gamma_{2}$ mode [Fig.~\ref{fig:cco_afm}(e)] involves shear modes along the $c$ direction associated with vibrations of the Ca and O atoms. Additionally, we find low-frequency optical modes arising from Cu and Ca vibrations, which contribute significantly to the total EPC [Fig.~\ref{fig:cco_afm}(b)]. Interestingly, none of the modes with strong EPC involve rotational phonons similar to those identified in the hole-doped AFM CCO.  When the EPC effects are included, the calculated spectral function of the VBM [Fig.~\ref{fig:cco_afm}(c)] shows pronounced band renormalization with substantial spectral broadening. These results strongly indicate that EPC plays an important role and that magnetic effects are essential for accurately capturing EPC effects in CCO. 

\section{EPC in C\MakeLowercase{a}C\MakeLowercase{u}O$_2$ with $G$-AFM configuration and 12.5\% hole doping}
\label{sec:epc_125}

\begin{figure}[htbp]
\centering
\includegraphics[width=\columnwidth]{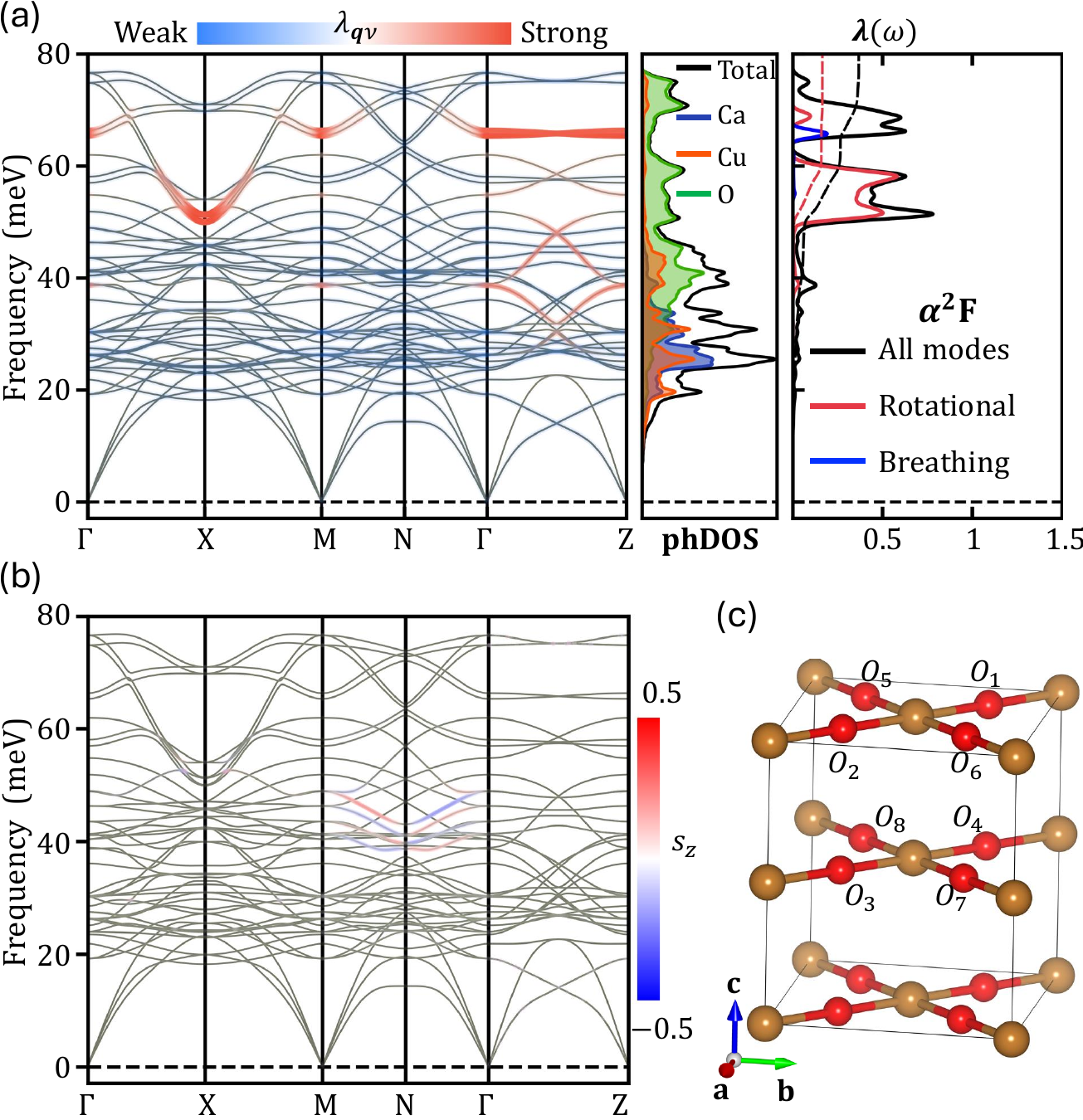}
\caption{\textbf{Phonons and EPC of $G$-AFM CaCuO$_2$ with 12.5\% hole doping.}
(a) Left: phonon dispersion along high-symmetry lines in the NM BZ; the marker color represents the mode-resolved EPC strength $\lambda_{\mathbf{q}\nu}$. Middle: total (black) and atom-projected (Ca, blue; Cu, orange; O, green) phonon density of states. Right: Eliashberg spectral function $\alpha^2F(\omega)$ (solid lines) and cumulative EPC strength $\lambda(\omega)$ (dashed lines) for all modes (black) and for the rotational oxygen modes (red); the solid blue line is $\alpha^2F(\omega)$ of the breathing modes.
(b) Phonon dispersion color-coded by the $z$ component of the oxygen circular polarization [Eq.~(1) of the main text].
(c) Crystal structure of the $G$-AFM configuration, with the Ca atoms omitted. Cu atoms are shown in brown and O atoms in red; the labels O$_1$--O$_8$ number the eight oxygen atoms.}
\label{fig:epc_125}
\end{figure}

Figure~\ref{fig:epc_125} shows the phonons and EPC of CCO in the $G$-AFM configuration at 12.5\% hole doping. As at 25\% doping, the strongly coupled modes with electrons are oxygen circular vibrations between about 50 and 70~meV, in particular the branch near 50~meV at $X$ and the flat branch near 66~meV along $\Gamma$--$Z$ [Fig.~\ref{fig:epc_125}(a)]. Integrating over all phonons yields $\lambda\approx0.36$, with $E_{\log}=52.7$~meV and $T_{\mathrm c}$ of 1.3--5.4~K for $\mu^*$ between 0.10 and 0.04 (Table~I of the main text). The rotational oxygen branches lie between about 40 and 52~meV near $X$ and along $M$--$N$--$\Gamma$ [Fig.~\ref{fig:epc_125}(b)], and the rotational modes dominate the two $\alpha^2F(\omega)$ peaks near 51 and 58~meV [red line in Fig.~\ref{fig:epc_125}(a)]. %

\section{Electronic structure of doped C\MakeLowercase{a}C\MakeLowercase{u}O$_2$ with $G$-AFM configuration}

\begin{figure}[htbp]
\centering
\includegraphics[width=\columnwidth]{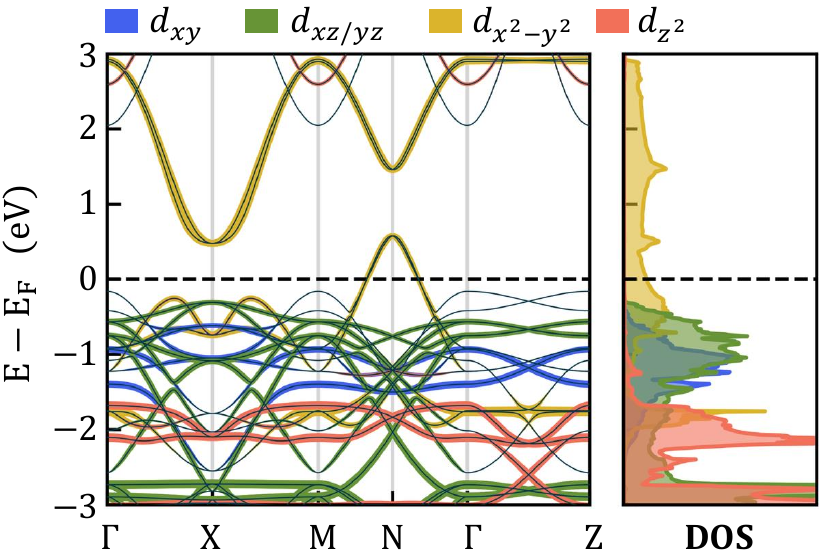}
\caption{\textbf{Calculated electronic structure of $G$-AFM phase of CaCuO$_2$ with 25\% hole doping.} Orbital-projected band structure along the high-symmetry lines in the BZ. Contributions of various Cu $d$-orbitals to the bands are identified; side panel shows the corresponding partial densities of states.}
\label{fig:doping_band}
\end{figure}

Figure~\ref{fig:doping_band} shows the calculated electronic structure of CCO in the $G$-AFM configuration at 25\% hole doping. The lower-Hubbard Cu 3d$_{x^2-y^2}$ band now crosses the Fermi level, so that our SCAN based calculations correctly capture the metal-insulator transition.

\section{Rotational modes and breathing modes in 25\% hole-doped C\MakeLowercase{a}C\MakeLowercase{u}O$_2$ with NM configuration}

\begin{figure*}[t!]
\centering
\includegraphics[width=\textwidth]{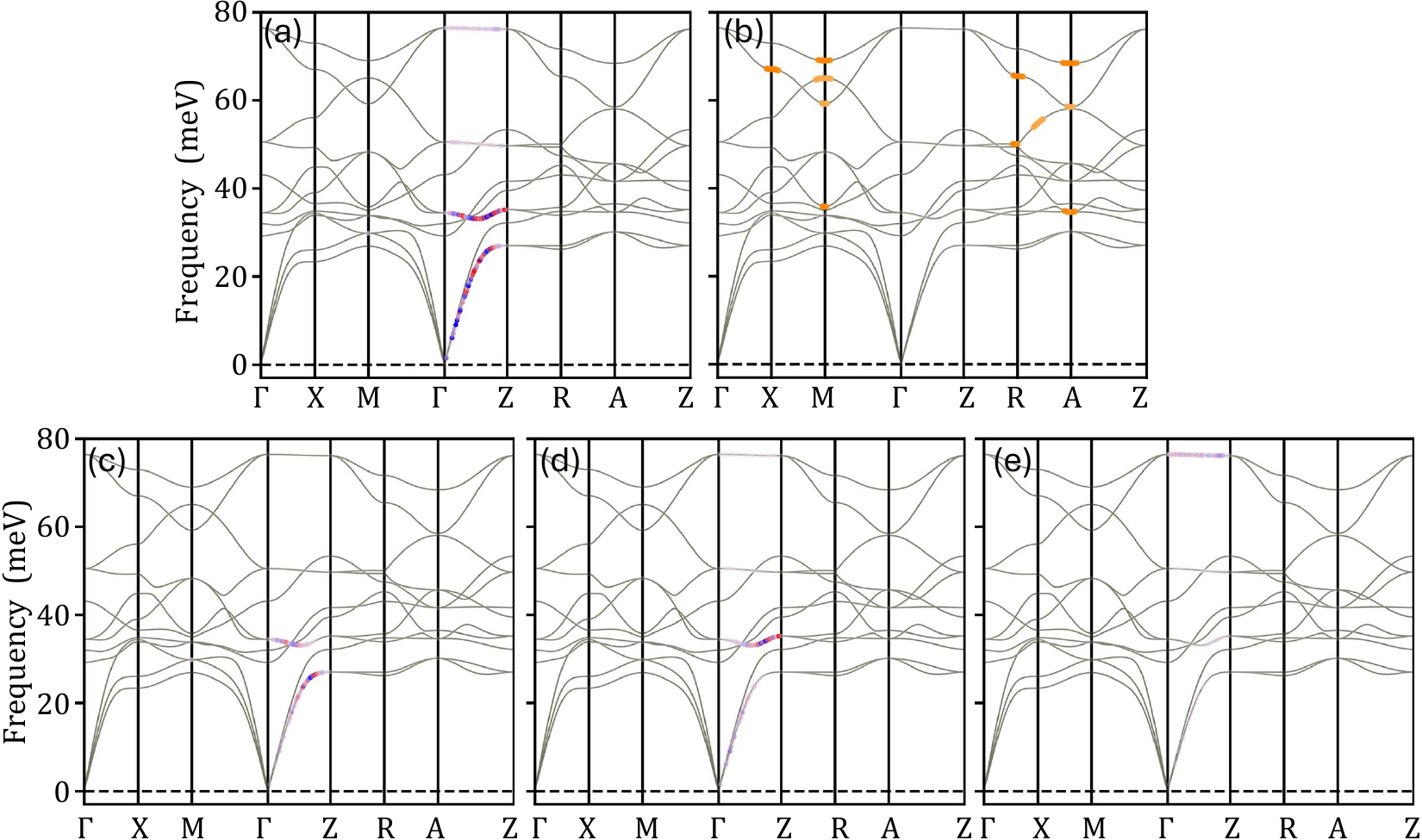}
\caption{\textbf{Phonon dispersion of 25\% hole-doped CaCuO$_2$ in the NM state, color-coded with the $z$-component of the phonon circular polarization}. 
(a) Phonon dispersions along high-symmetry lines, showing the total $z$-component of the angular momentum summed over all atoms. Red, blue, and gray colors denote right-handed ($s^{z}_{\mathbf{q},\mu}>0$), left-handed ($s^{z}_{\mathbf{q},\mu}<0$), and unpolarized ($s^{z}_{\mathbf{q},\mu}=0$) modes, respectively. 
(b) Same as (a) except that breathing modes are highlighted in yellow color. 
(c) Same as (a), except that the $z$-component of the angular momentum projected onto Ca atoms is shown with the same color scheme as in (a). 
(d) Same as (a), except that the projection onto Cu atoms is shown. 
(e) Same as (a), except that the projection onto one O atom is shown.}
\label{fig:nm_vibproj}
\end{figure*}

In the NM phase of 25$\%$ hole-doped CCO, we analyze phonon circular polarization using Eq.~(1) of the main text. Figure~\ref{fig:nm_vibproj}(a) shows phonon dispersions along the high-symmetry lines in The BZ, color-coded with the total $z$-component of the angular momentum summed over all atoms. The atom-resolved projections in Figs.~\ref{fig:nm_vibproj}(c)--(e) indicate that there are degenerate phonon modes which can be expressed in a chiral basis in the NM case mainly arising from low-frequency branches, with contributions roughly split between Ca and Cu motions ($\approx55\%$ for Ca vs. 40$\%$ for Cu) and negligible involvement of oxygens. Fig.~\ref{fig:nm_vibproj}(b) focuses on the breathing-like modes, which remain largely non-rotational in the NM state. The NM phase thus shows no oxygen-dominated rotational modes of the type that are found in the AFM case [Fig.~\ref{fig:nm_vibproj}(e)], although there is a degenerate pair of bands in the highest acoustic branch along $\Gamma-Z$ that can become chiral with appropriate symmetry breaking, but with negligible EPC strength.

\section{Rotational modes and breathing modes in $G$-AFM configuration of C\MakeLowercase{a}C\MakeLowercase{u}O$_2$ with 25\% hole doping}
\label{sec:rot_afm}

\begin{figure*}[t!]
\centering
\includegraphics[width=\textwidth]{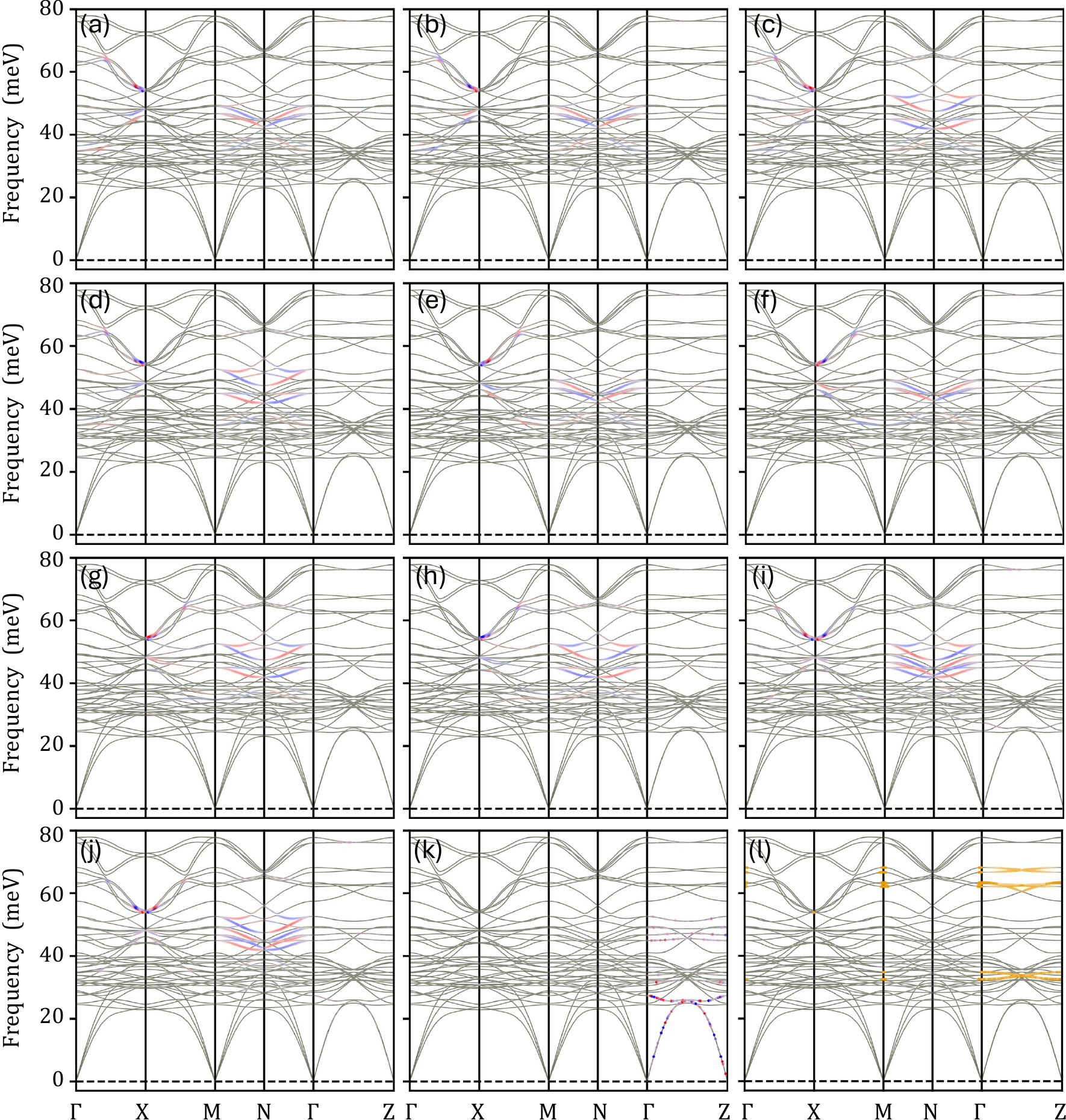}
\caption{\textbf{Phonon dispersion of hole-doped CaCuO$_2$ in the $G$-AFM phase, color-coded by the $z$ component of the phonon circular polarization}. 
(a)--(h) Phonon dispersions along the high-symmetry lines in the BZ, showing the site-resolved $z$ component of the angular momentum projected onto each oxygen atom [O$_1$--O$_8$; subscripts label the oxygen sites in the $G$-AFM unit cell, as numbered in Fig.~\ref{fig:epc_125}(c)]. Red, blue, and gray colors denote right-handed ($s^{z}_{\mathbf{q},\sigma}>0$), left-handed ($s^{z}_{\mathbf{q},\sigma}<0$), and unpolarized ($s^{z}_{\mathbf{q},\sigma}=0$) modes, respectively. 
(i) Same as panels (a)--(h), except that the summed projection $s^{z}$(O$_1$)+$s^{z}$(O$_3$)+$s^{z}$(O$_6$)+$s^{z}$(O$_8$) is shown. 
(j) Same as (i), except that the complementary sum $s^{z}$(O$_2$)+$s^{z}$(O$_4$)+$s^{z}$(O$_5$)+$s^{z}$(O$_7$) is shown. 
(k) Same as (i), except that the results of a summation over all atoms in the unit cell are shown. 
(l) Phonon dispersion with oxygen breathing-like modes highlighted in yellow.}

\label{fig:rotational_proj}
\end{figure*}

There are eight oxygen atoms in the $G$-AFM unit cell of CCO, which we number as illustrated in Fig.~\ref{fig:epc_125}(c). Using Eq.~(1) of the main text, we compute the $z$ component of the angular momentum for each oxygen atom. Figures~\ref{fig:rotational_proj}(a)--(h) show the phonon dispersions along the high-symmetry lines, color-coded by the site-resolved $z$-polarization projected onto each of the eight oxygen atoms. These plots reveal a rich set of rotational oxygen vibrations in the $G$-AFM phase. Because an equal number of oxygen atoms rotate clockwise and counterclockwise, the net oxygen circular polarization obtained by summing over all oxygen atoms vanishes for the rotational modes [Fig.~\ref{fig:rotational_proj}(k)], although degenerate chiral pairs along the $\Gamma\!-\!Z$ direction [Fig.~\ref{fig:rotational_proj}(k)] can be observed in the low-frequency transverse acoustic branches dominated by Ca motion, which is also the case in the undoped NM phase [Fig.~\ref{fig:nm_vibproj}(c)]. 

We note that the spotty non-continuous red and blue colors along the transverse acoustic modes in Fig.~\ref{fig:nm_vibproj}(a), (c), and (d) seemingly indicate a highly oscillatory momentum dependent direction of the phonon angular momentum $s^{z}_{\mathbf{q},\mu}$. However, since the transverse acoustic phonon modes are degenerate their angular momentum character at each $\mathbf{q}$ is mixed, thereby yielding a non-continuous result. This stems from the numerical diagonalization routine not knowing how to appropriately assign angular momentum character to the states in the degenerate subspace. This can be corrected by introducing an infinitesimal symmetry breaking field that puts the angular momentum states in the correct order along the momentum path in the Brillouin zone. As discussed in the main text, however, these acoustic modes exhibit only weak EPC. 

\FloatBarrier
\section{Rotational modes in a paramagnetic $2\times2\times2$ SQS supercell with 25\% hole doping}

\begin{figure}[htbp]
\centering
\includegraphics[width=\columnwidth]{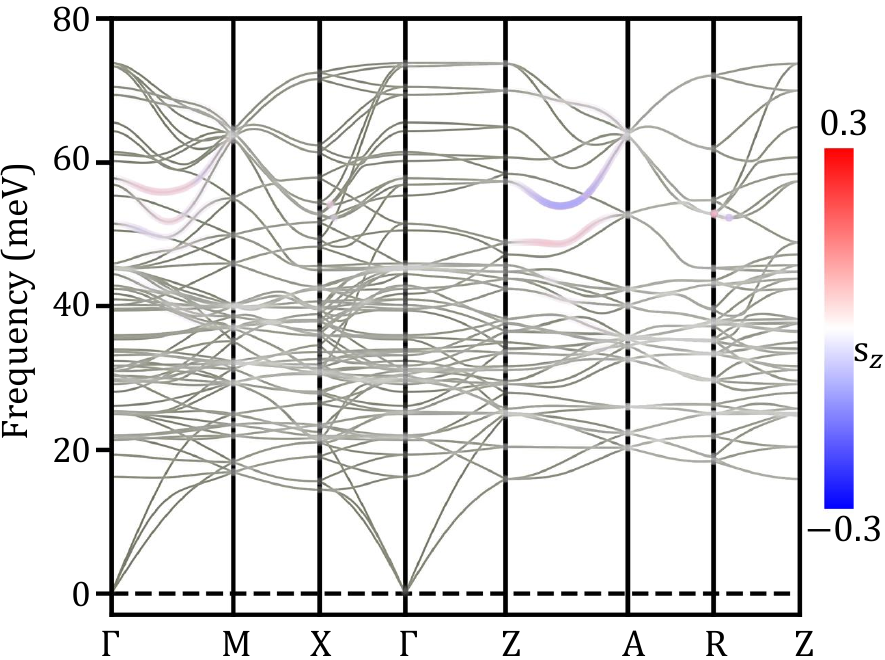}
\caption{\textbf{Phonon dispersion of the paramagnetic $2\times2\times2$ SQS supercell of CaCuO$_2$ with 25\% hole doping}, color-coded by the $z$ component of the circular polarization of the oxygen atom O$_{14}$. Red and blue denote right-handed ($s^{z}>0$) and left-handed ($s^{z}<0$) modes, respectively. $\Gamma$, $M$, $X$, $Z$, $A$, and $R$ are the high-symmetry points of the supercell BZ.}
\label{fig:sqs}
\end{figure}

To test whether the rotational oxygen modes also appear in other magnetic configurations, we calculated the phonon dispersion of a paramagnetic $2\times2\times2$ special quasirandom structure (SQS) supercell of CCO at 25\% hole doping.
Figure~\ref{fig:sqs} shows the dispersion, color-coded by the $z$ component of the circular polarization of the oxygen atom O$_{14}$, i.e., the O$_{14}$ term of Eq.~(1) of the main text. Rotational modes appear between about 45 and 62~meV along $\Gamma$--$M$--$X$ and $Z$--$A$--$R$, in the same energy range as the rotational oxygen branches of the $G$-AFM cell (Sec.~\ref{sec:rot_afm} and Fig.~2 of the main text). The rotational oxygen modes therefore also appear in a paramagnetic configuration with disordered Cu magnetic moments, without long-range $G$-AFM order. The EPC of the SQS supercell was not calculated because of its size.

\section{Dependence of the rotational-mode contribution on the classification threshold}
\label{sec:threshold}

\begin{figure}[htbp]
\centering
\includegraphics[width=\columnwidth]{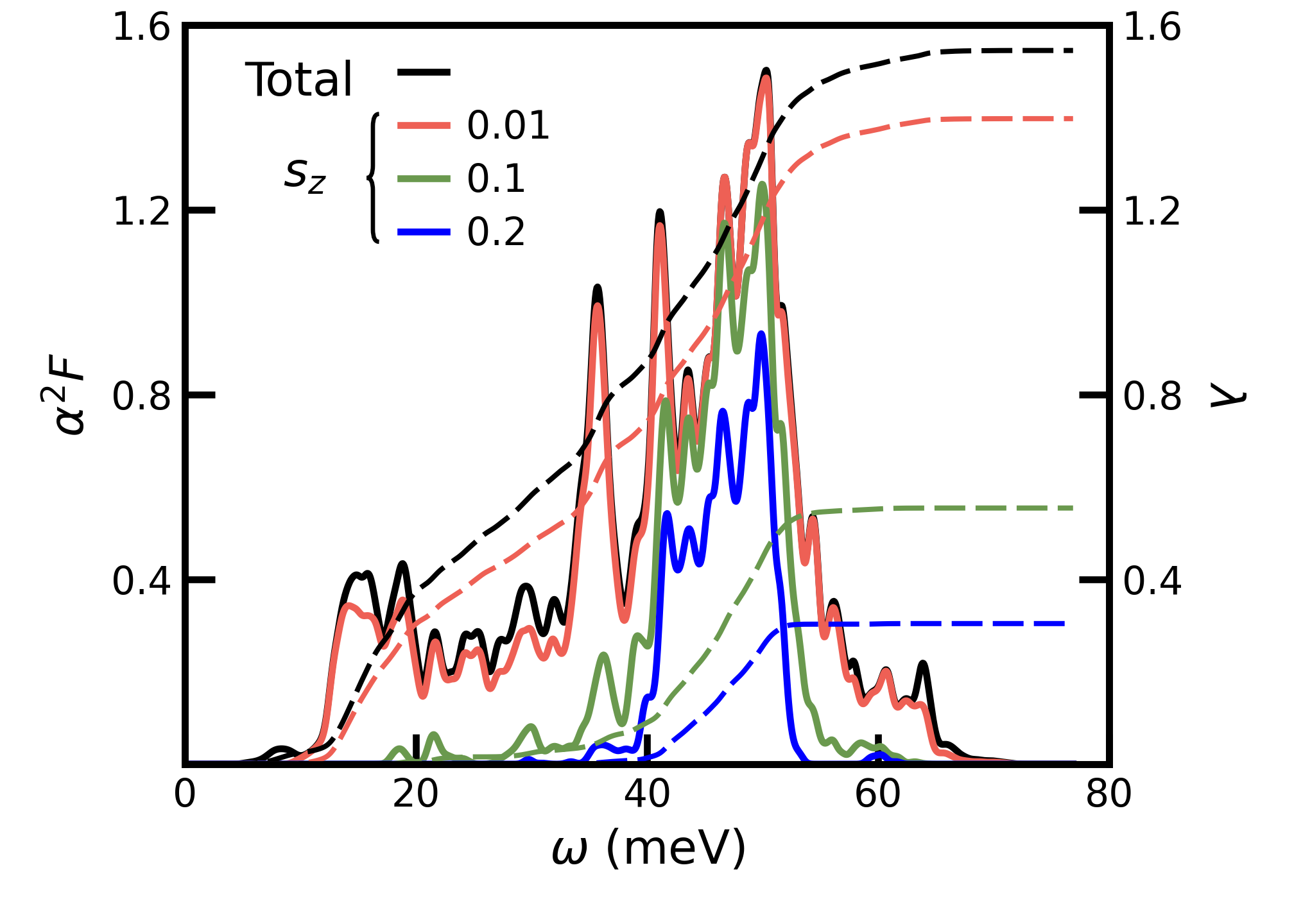}
\caption{\textbf{Dependence of the rotational-mode decomposition on the classification threshold for $G$-AFM CaCuO$_2$ with 37.5\% hole doping.} Eliashberg spectral function $\alpha^2F(\omega)$ (solid lines, left axis) and cumulative EPC strength $\lambda(\omega)$ (dashed lines, right axis) for all modes (black) and for the rotational modes with $s_{\mathrm c}=0.01$ (red), 0.1 (green), and 0.2 (blue). A mode is classified as rotational if $|\mathbf{e}^{\dagger}_{i,\mathbf{q},\mu} S^{z}\mathbf{e}_{i,\mathbf{q},\mu}|\geq s_{\mathrm c}$ for at least one O site $i$. }
\label{fig:a2f-37.5}
\end{figure}

In the main text, a phonon mode $(\mathbf q,\mu)$ is classified as rotational if $|\mathbf{e}^{\dagger}_{i,\mathbf{q},\mu} S^{z}\mathbf{e}_{i,\mathbf{q},\mu}|\geq s_{\mathrm c}$ for at least one O site $i$, with $s_{\mathrm c}=0.1$ [Eq.~(1) of the main text]. The rotational contributions to $\alpha^2F(\omega)$ and $\lambda$ are obtained by summing the mode-resolved contributions over these modes. Each mode is assigned as a whole at its own $\mathbf q$ point, so no branch tracking is needed, and the classification does not depend on the phase of the eigenvector. To test how the decomposition depends on the threshold, Fig.~\ref{fig:a2f-37.5} compares the rotational contributions for $s_{\mathrm c}=0.01$, 0.1, and 0.2 at 37.5\% hole doping.

For $s_{\mathrm c}=0.01$, nearly all modes are classified as rotational: the rotational $\alpha^2F(\omega)$ almost coincides with the total one over the whole spectrum, including the low-frequency modes between 10 and 35~meV, and $\lambda_{\mathrm{rot}}\approx1.40$, compared with the total $\lambda\approx1.57$. This threshold therefore also counts nearly linear oxygen vibrations with only a small circular component, such as the mode $C_2$ at 12.3~meV, which combines linear Cu displacements with elliptical O motions (Fig.~5 of the main text). For $s_{\mathrm c}=0.1$ and 0.2, the rotational contribution is confined to about 35--55 and 40--53~meV, with $\lambda_{\mathrm{rot}}\approx0.56$ and 0.31, respectively. For both thresholds, the rotational modes dominate the main $\alpha^2F(\omega)$ peak near 50~meV. The finding that rotational oxygen modes dominate the coupling near 50~meV is thus insensitive to the choice of $s_{\mathrm c}$ between 0.1 and 0.2, whereas the value of $\lambda_{\mathrm{rot}}$ depends on $s_{\mathrm c}$.  Similar results are found for 25\% dopings; in the absence of any mode-dependent method of separating rotational and non-rotational contributions to $\lambda$, we make the simple approximation os using $s_0=0.1$ to determine whether a mode is rotational or not.

\FloatBarrier
\section{CDW-related phonons}

The CDW phase of cuprates presents a further avenue for validating our computational framework. Note that the conventional AFM order is associated with hot spots, defined as points on the Fermi surface that cross the AFM BZ boundary. Long-range AFM order gaps the Fermi surface at these hot spots, breaking it into a necklace of alternating electron and hole pockets with doping-dependent areas, as observed in electron-doped cuprates~\cite{NCCO1,NCCO2}.  Hole-doped cuprates exhibit more complex orders with doping~\cite{Tranquada1996,CDW1,CDW2}. Charge stripes emerge at low doping levels, with a $q$-vector that increases with doping, inconsistent with Fermi surface nesting but consistent with the tendency of charge stripes to maximize their spatial separation. At higher doping levels ($x>1/8$), a CDW appears with a $q$-vector that decreases with doping, consistent with possible Fermi surface nesting. The identity of the nested Fermi surface, however, remains unclear. Nesting of the NM Fermi surface is inconsistent with the observed $Q_{CDW}$, and while hot-spot nesting comes closer, it is not quantitatively accurate. There is evidence, however, that small electron pockets exist that are sufficiently well ordered to show quantum oscillations and cross over to a large Fermi surface at the pseudogap collapse; moreover, $Q_{CDW}$ is found to be consistent with the inter-electron-pocket nesting vector closest to hot-spot nesting~\cite{Paper2}. 

\begin{figure}[htbp]
\centering
\includegraphics[width=\columnwidth]{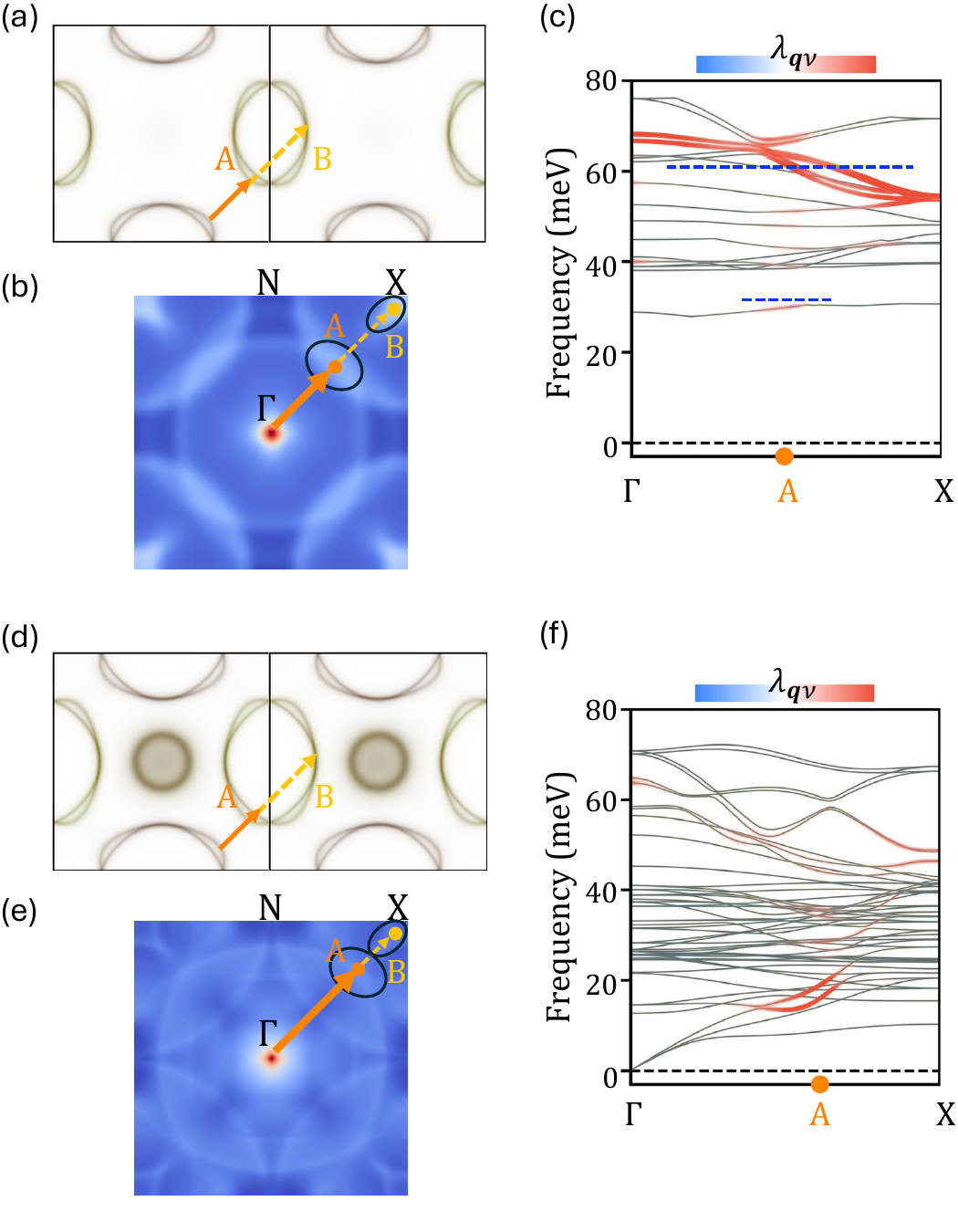}
\caption{\textbf{CDW related phonons in C\MakeLowercase{a}C\MakeLowercase{u}O$_2$}. (a-c) 25\% doping.
(a) AFM Fermi surfaces in $\Gamma$ plane of the AFM BZ. Colored arrows mark the nesting vectors along the NM BZ $\Gamma-X$ line. (b) Nesting map in AFM BZ. $\Gamma$ lies at the center of the map, $N$ at the top edge, and $X$ at the top-right corner. The orange arrow points from $\Gamma$ to the nesting peak A, and the dashed yellow arrow extends it to the peak B along $\Gamma$--$X$. Peak A lies at $\mathbf q=(0.25,0)$, in reciprocal-lattice units of the nonmagnetic cell, and corresponds to $Q_{\mathrm{CDW}}$; the maximum values at A and B are 4.43 and 6.34.
(c) Blowup of the $\Gamma$--$X$ phonon dispersion showing the strength of EPC, taken from Main Fig.~1(a). The solid orange dot indicates the average nesting vector from panel (a), and the two horizontal blue dashed lines indicate experimental phonon modes of a Bi-cuprate with large linewidth broadening due to strong EPC~\cite{Ding_PNAS}. (d-f) Similar results for 37.5\% doping. In (e), peak A lies at $\mathbf q=(0.31,0)$, and the maximum values at A and B are 11.11 and 9.53. While there is also nesting of the $\Gamma$-pocket in (d) (purple arrows), in (e) we focus on the $d_{x^2-y^2}$ nesting.}

\label{fig:S6}
\end{figure}
\begin{table*}[htbp]
\centering
\caption{Numerical settings for the thermal-Hall calculation.}
\label{tab:settings}
\begin{tabular}{ll}
\toprule
Setting & Value \\
\midrule
Force-constant model & DFT HR (BNS 123.347) \\
Reciprocal-space mesh & $51\times51\times51$ with a half-grid shift \\
Field scan & 51 equally spaced values of $r$ from 0 to 0.2 at 50, 150, and 300 K \\
Temperature scan & 5--300 K in 5 K increments at $r=0.2$ \\
Gyroscopic atoms & Cu sites 5--8 \\
Staggered signs & $(1,1,-1,-1)$ \\
Gyroscopic scale & $\lambda_{\mathrm{AF}}=\SI{1}{THz}$ \\
Uniform-field range & $0\leq r\leq0.2$, corresponding to
$0\leq B_{\mathrm{eff}}\leq\SI{7.145}{T}$ \\
Gyroscopic axis & Crystallographic $z$ direction \\
Minimum retained frequency & $10^{-5}$ THz \\
Near-degeneracy tolerance & $10^{-4}$ THz \\
Effective bulk thickness & \SI{6.56179576618}{\angstrom} \\
\bottomrule
\end{tabular}
\end{table*}

In this context, it is natural to ask whether our computational framework can predict the correct Fermi surface nesting vector and reproduce the associated phonon softening. Experimental studies suggest that these signatures are highly unusual: the atomic displacement pattern in the CDW state indicates that the soft mode is an {\it optical (breathing mode) phonon}~\cite{Forgan2015_natcom}, and in a Bi-cuprate at least three phonon branches show signs of softening near the same $q$-vector~\cite{Ding_PNAS}. These results are suggestive of a Kohn anomaly modified by anticrossing effects involving several phonon modes, constituting a Kohn ladder~\cite{NbSe2_Susy}.

Our findings are consistent with this picture. Fig.~\ref{fig:S6}(a) shows the AFM Fermi surfaces at 25\% doping plotted in the AFM BZ, highlighting two possible nesting vectors: A,  the near-hot-spot interpocket nesting vector (orange arrow) and B, an alternate inter-pocket nesting vector (yellow arrow); note that the arrows sum up to $\Gamma\rightarrow X$. The corresponding nesting function is shown in Fig.~\ref{fig:S6}(b), where the arrows have the same lengths as the corresponding arrows in (a). Note that $q_A$ and $q_B$ are both approximately $(\Gamma\rightarrow X)/2=(0.255,0)$ [orange dot in Fig.~\ref{fig:S6}(c)], which is in good agreement with the experimental $Q_{CDW}=(0.25,0)$.

Finally, Fig.~\ref{fig:S6}(c) compares our EPC map from Main Fig.~1(a), restricted to branches with strong EPC between $\Gamma$ and $X$, with the experimental results of Ref.~\cite{Ding_PNAS}. In addition to evidence for a CDW at $Q_{CDW}=(0.25,0)$, the experiments found two phonon bands with significant EPC, as adduced from their broad linewidths, indicated by the two horizontal blue dashed lines at 30 and 60 meV, identified as bond-bending and oxygen breathing modes, respectively~\cite{Ding_PNAS}.  While our theoretical modes cluster into two bands in reasonable proximity to experiment, the upper branches are associated with oxygen rotational modes and for the lower branch we find a stronger EPC associated with a {\it Cu breathing mode} at 44~meV, which is also an SSH phonon, in that the Cu-breathing motion directly modulates the Cu-Cu distance and hence the hopping $t$. In contrast, the bond-bending mode proposed in Ref.~\cite{Ding_PNAS} is found at 38~meV and produces a somewhat weaker peak in $\alpha^2F$. We note, however, that our doping is significantly higher than the experimental doping, and we find no evidence of the very low-energy CDW-related features observed experimentally.

Frames (d-f) repeat the same results for the 37.5\% doped sample.  We note in particular that frame (f) shows a very similar Kohn ladder, but with considerable phonon softening.  In particular, the mode $C_2$ at 12.3 meV shows a dip in the dispersion associated with strong EPC, while all three acoustic modes show rotational Ca atoms with softening, although neither rotational oxygen nor significant contributions to $\lambda$.  The picture is further complicated since most of the phonon spectra display softening, in particular along the remaining acoustic branches,  $C_2$ and the acoustic modes all have maximal softening along $\Gamma\rightarrow X$, albeit rather closer to $(\Gamma\rightarrow X)/2$ than expected from the AFM nesting map, frames (d) and (e).

\FloatBarrier
\section{Thermal Hall effect}
\begin{figure*}[t!]
 \centering
 \includegraphics[width=.9\textwidth]{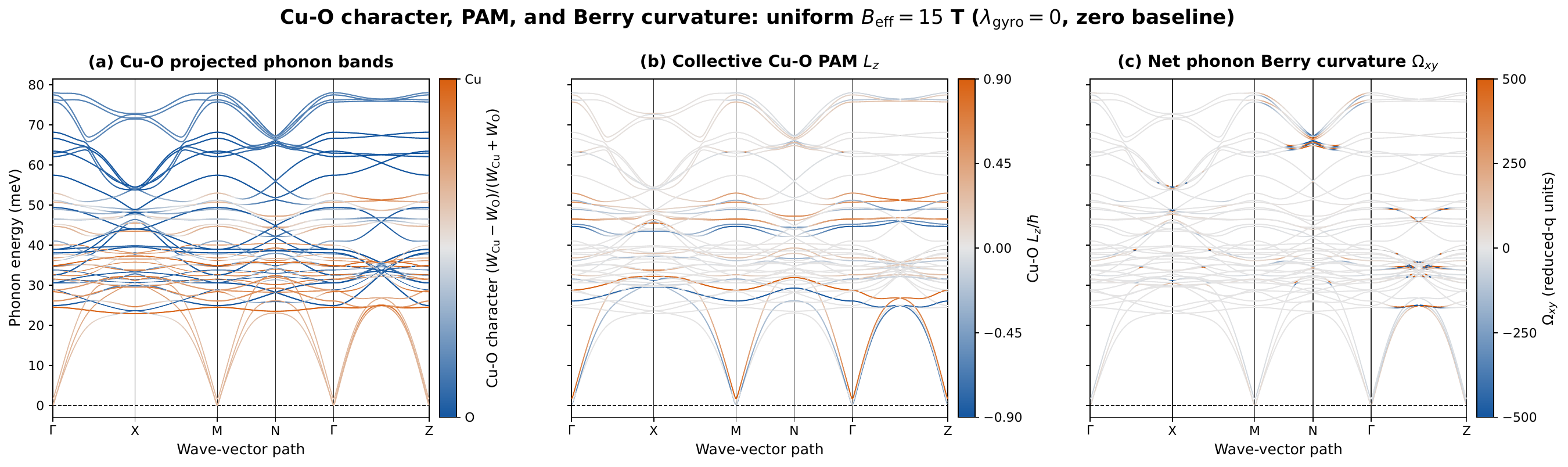}
 \caption{Cu--O character, collective PAM, and phonon Berry curvature along the selected high-symmetry path for $B_{eff} = 15$T.  The first panel identifies the Cu and O weights; the second shows that the angular momentum belongs to collective Cu--O eigenmodes; and the third displays the net mode-resolved Berry curvature generated after time-reversal symmetry is broken.}
 \label{fig:path}
\end{figure*}
In this section, we will discuss the role of predicted rotational phonons of oxygen atoms in achieving the thermal Hall effect. The circular motion of atoms in rotational phonons carries a real angular momentum \cite{Juraschek2019PhononMoments}. Due to this angular momentum, the eigenvectors of the rotational phonon acquire a momentum-dependent geometric phase and curvature when subjected to a perturbation like a magnetic field or spin-phonon coupling. The resulting Berry curvature is concentrated near local phonon-band extrema and allowed crossings involving modes with strong circular polarization. A finite intrinsic thermal Hall conductivity requires a nonzero thermally weighted Brillouin-zone integral of the phonon Berry curvature. In the pristine structure, with time-reversal symmetry intact, symmetry-related oxygen atoms at $q$ and $-q$ will carry opposite local angular momenta.  These staggered local contributions cancel and do not by themselves generate a macroscopic phonon thermal Hall response \cite{Qin2012PhononHall}.
For a nonmagnetic system that preserves time-reversal symmetry,
\begin{equation}
\mathcal{T}: \Omega_{n, x y}(\mathbf{q}) \rightarrow-\Omega_{\tilde{n}, x y}(-\mathbf{q})
\end{equation}
so the contributions from opposite wave vectors cancel.  If inversion symmetry is also present, $\mathcal{P}: \Omega_{n, x y}(\mathbf{q}) \rightarrow \Omega_{\tilde{n}, x y}(-\mathbf{q})$ and the Berry curvature vanishes pointwise for an isolated nondegenerate band.  An external magnetic field supplies one possible time-reversal-symmetry breaking. 

\subsection{Effective model}
To realize this, we have used an effective lattice model, which introduces a gyrotropic velocity-force coupling.  This term distinguishes opposite handedness, lifts their cancellation, and permits the coupled Cu-O bands to acquire a nonzero net Berry curvature. Starting from the mass-weighted dynamical matrix $D(\mathbf q)$, the time-reversal-breaking phonon eigenvalue problem is written as
\begin{equation}
 \left[D(\mathbf q)-\omega_{n\mathbf q}^{2}I
 -2i\omega_{n\mathbf q}G{(r)}\right]
 \mathbf e_{n\mathbf q}=0,
 \label{eq:gyro_SM}
\end{equation}
where,
$$
G(r)=r J_z,~J_z=\left(\begin{array}{ccc}
0 & -1 & 0 \\
1 & 0 & 0 \\
0 & 0 & 0
\end{array}\right).
$$
Because the Cu and O displacements are dynamically coupled, the eigenvectors describe collective Cu--O motion rather than a Cu-only or O-only response (see Fig.~\ref{fig:path}(a)). The calculations use $r$ as an imbalance parameter that breaks time-reversal symmetry and simulates a uniform magnetic field. The $r=0$ point is the compensated staggered model, while a finite value of $r$ removes the compensation. We define the corresponding Zeeman-equivalent effective magnetic field in terms of $r$ as follows:
\begin{equation}
B_{\mathrm{eff}}=\frac{h}{g_{\mathrm{eff}}\mu_B}\,r,
\label{eq:field_conversion}
\end{equation}
where $h$ is Planck's constant and $\mu_B$ is the Bohr magneton.
Using $g_{\mathrm{eff}}=2$ and expressing $r$ in THz gives
\begin{equation}
\begin{gathered}
B_{\mathrm{eff}} = 35.7239 r~\,\frac{\mathrm{T}}{\mathrm{THz}}\,\\
r=0.419887~\mathrm{THz}
\Longleftrightarrow B_{\mathrm{eff}}=15~\mathrm{T}.
\end{gathered}
\end{equation}
This conversion is only an approximate phenomenological calibration. A quantitative calculation of the magnetic-field-induced ionic Lorentz force would require the relevant Born effective-charge tensors. 

\subsection{Phonon angular momentum and Berry curvature}

For the mass-normalized polarization vector of atom $a$ in mode $(n,\mathbf q)$, the out-of-plane phonon angular momentum (PAM) is
\begin{equation}
 \frac{L^z_{a n}(\mathbf q)}{\hbar}
 =2\operatorname{Im}\!\left[e^{*}_{a n x}(\mathbf q)e_{a n y}(\mathbf q)\right].
 \label{eq:pam}
\end{equation}
The total PAM is obtained by summing over a complete set of atoms as represented in Fig.~\ref{fig:path}(b).  An atom-resolved or non-closed subset can be finite even when the complete symmetry orbit cancels.

For an isolated positive-frequency phonon band, the Berry curvature is obtained from the left and right eigenvectors of the first-order phase-space problem (see Fig.~\ref{fig:path}(c)).  At exact or near degeneracies, individual-band curvature is gauge dependent; the physically robust quantity is the trace of the non-Abelian curvature over the complete nearly degenerate manifold.  All thermal-Hall integrations reported below use a half-shifted mesh to avoid direct evaluation at the acoustic touching at $\Gamma$.

\subsection{Intrinsic phonon thermal Hall conductivity}

\begin{figure}[htbp]
 \centering
 \includegraphics[width=\columnwidth]{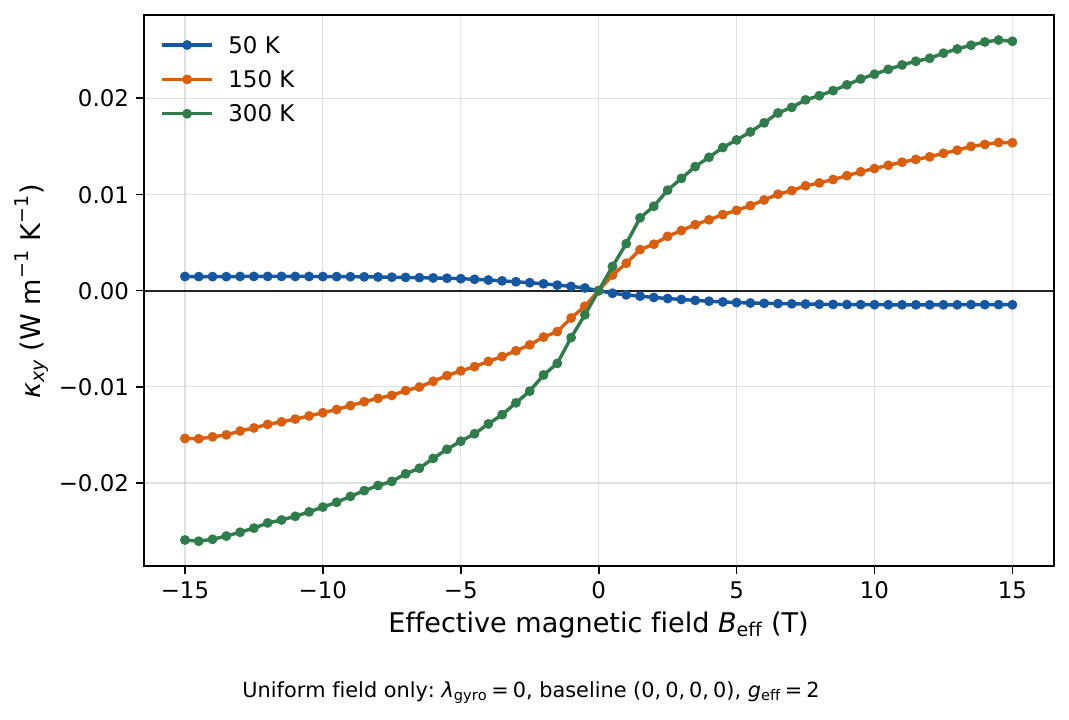}
 \caption{Intrinsic phonon thermal Hall conductivity as a function of external magnetic field obtained from the DFT force-constant model on the $101^3$ shifted mesh and $r=0.42$ THz. }
 \label{fig:kappa_field}
\end{figure}
\begin{figure}[htbp]
 \centering
 \includegraphics[width=\columnwidth]{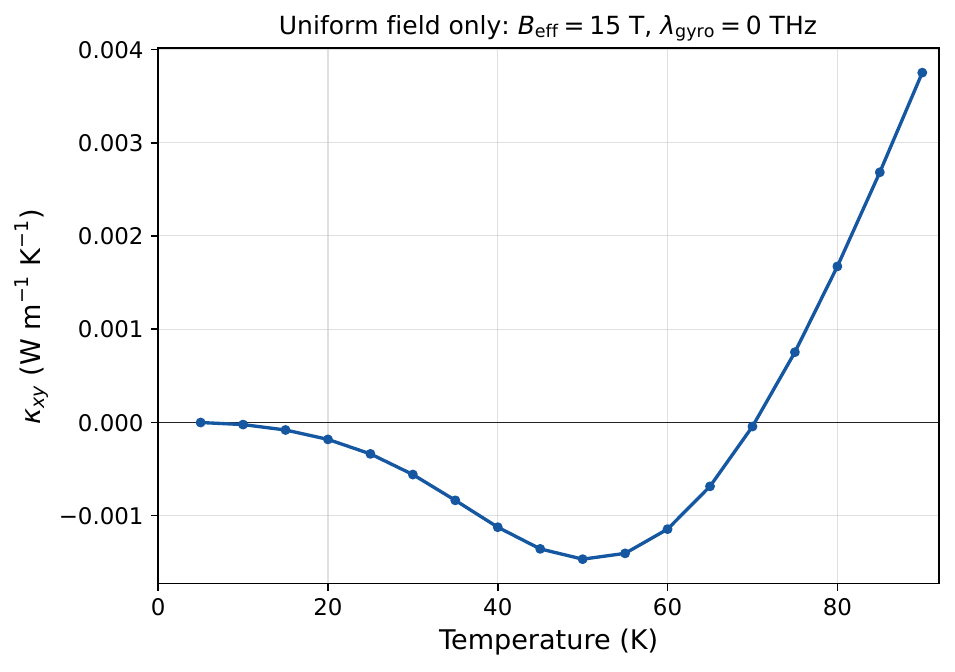}
 \caption{Temperature dependence of the intrinsic phonon thermal Hall conductivity for the   DFT force-constant model at fixed $r=0.42$ THz, corresponding to $B_{\mathrm{eff}}=\SI{15}{T}$.  The calculation uses the same $101^3$ shifted mesh as Fig.~\ref{fig:kappa_field}.}
 \label{fig:kappa_temperature}
\end{figure}

The intrinsic phonon thermal Hall conductivity is evaluated as
\begin{equation}
 \kappa_{xy}
 =-\frac{k_B^2T}{\hbar}
 \sum_{n>0}\int_{\mathrm{BZ}}\frac{d^3q}{(2\pi)^3}
 c_2\!\left[n_B(\omega_{n\mathbf q})\right]
 \Omega_{n,xy}(\mathbf q),
 \label{eq:kappa}
\end{equation}
where $n_B$ is the Bose--Einstein distribution and
\begin{equation}
 c_2(\rho)=(1+\rho)
 \left[\ln\!\left(\frac{1+\rho}{\rho}\right)\right]^2
 -(\ln\rho)^2-2\operatorname{Li}_2(-\rho),
\end{equation}
where $\mathrm{Li}_2$ represents the dilogarithm.  The frequency-dependent Bose weighting means that $\kappa_{xy}$ need not follow the unweighted sum of Berry curvature band by band.  

Figure~\ref{fig:kappa_field} shows the calculated phonon thermal Hall conductivity as a function of $B_{\mathrm{eff}}$ at several temperatures. The response is approximately odd in $B_{\mathrm{eff}}$ and exhibits a nonlinear field dependence. For positive fields, $\kappa_{xy}$ is negative at 50~K but positive at 150 and 300~K. At $B_{\mathrm{eff}}=15$~T, the corresponding values are approximately $-1.47\times10^{-3}$~\si{W\,m^{-1}\,K^{-1}}, $1.54\times10^{-2}$~\si{W\,m^{-1}\,K^{-1}}, and $2.59\times10^{-2}$~\si{W\,m^{-1}\,K^{-1}}, respectively. This nonmonotonic temperature dependence arises because the field-like gyroscopic term shifts and gaps the crossings that carry the dominant Berry curvature.

The temperature dependence at $B_{\mathrm{eff}}=15$~T is shown in Fig.~\ref{fig:kappa_temperature}. Over the calculated 5--90~K range, $\kappa_{xy}$ reaches a minimum near 50~K, changes sign between 70 and 75~K, and increases to approximately $3.75\times10^{-3}$~\si{W\,m^{-1}\,K^{-1}} at 90~K. Within the present harmonic model, temperature enters through the
occupation-dependent weighting of the phonon Berry curvature. The sign reversal is therefore consistent with a changing balance between contributions of opposite sign. Main-text Fig.~7(c) shows the corresponding $\kappa_{xy}/T$
for comparison with experiment.

Comparison with experiment indicates that the applied magnetic field alone is unlikely to account for the calculated thermal Hall response, suggesting that additional time-reversal-symmetry-breaking mechanisms must contribute -- i.e., $B_{eff}$ should be $\gg$ 15T.

\FloatBarrier

\end{document}